\documentclass[a4paper,11pt]{article}
\usepackage[utf8]{inputenc}
\usepackage{hyperref}
\usepackage{amsfonts}
\usepackage{amsmath}
\usepackage{amssymb}
\usepackage{indentfirst}
\usepackage{graphicx}
\usepackage{cite}
\usepackage{color}
\usepackage{xcolor}

\usepackage{breqn}

\usepackage{setspace}%

\hypersetup{
colorlinks=true,
citecolor=black,
linkcolor=black,
filecolor=black,
urlcolor=black,
}
\numberwithin{equation}{section}
\allowdisplaybreaks
\makeatletter

\begin{document}

\begin{titlepage}

\rightline{  }

\baselineskip=24pt

\begin{center}
\textbf{\LARGE{Non-linear realizations and invariant action\\ principles in higher
gauge theory}}
\par\end{center}{\LARGE \par}

\begin{center}
	\vspace{0.5cm}
	S. Salgado
	\small
	\\[4mm]
	\textit{Instituto de Alta Investigaci\'on, Universidad de Tarapac\'a, Casilla 7D,
Arica, Chile}\\[0.2mm]
\vskip 7pt
	\footnotesize
	\texttt{E-mail: sebasalg@gmail.com}
	 \\[3mm]
 
	\par\end{center}
\vskip 20pt
\begin{abstract}
\noindent We propose an extension of the formalism of non-linear realizations to the
case of FDAs. We first consider the case of FDAs carrying one $p$-form
extension and no non-trivial cohomology. We show that it is possible to define
large gauge transformations as a direct extension of those induced by their
Lie subalgebras, and study the resulting non-linear realizations. Furthermore,
we extend the results to the case FDAs with non-trivial cohomology by
introducing large gauge transformations that carry the information about the
FDA cocycle structure constants. We consider two bosonic examples of this type
of gauge algebra, namely, FDA extensions of the Poincar\'{e} and Maxwell
algebras, write down their dual $L_{\infty}$ algebras and study their
non-linear realizations and possible invariant action principles. Finally,
consider a similar treatment on the FDA of $D=11$ supergravity, by deriving
its dual algebra and presenting a non-linear realization that allows a gauge
invariant formulation of its action principle.

	\end{abstract}
\end{titlepage}\newpage {} 

\noindent\rule{162mm}{0.4pt}
\tableofcontents
\noindent\rule{162mm}{0.4pt}

\section{Introduction}

The use of non-linear realizations of Lie algebras allows the construction of
classical gauge theories genuinely invariant under a gauge group
\cite{Stelle:1979aj,Ivanov:1981wn,Ivanov:1981wm,Grignani:1991nj}. Examples of
this can be found in the context of gravity and supergravity theories,
particularly for the cases of the Poincar\'{e} and AdS groups and supergroups
\cite{Salgado:2001bn,Salgado:2003rf}. Such formalism, sometimes reffered as
Stelle--West Grignani-Nardelli (SW-GN) formalism, involves separating the
gauge group in a stability subgroup, usually the Lorentz group, and the
remaining subspace. The corresponding Lie group, typically chosen as the
Poincar\'{e} or AdS group, is gauged by introducing a one-form gauge
connection with non-vanishing curvature. In gravity theories, when gauging a
Lie group, the vielbein field encoding the information about the spacetime
metricity is usually identified as the gauge field associated with the
translation generators. However, in the SW-GN formalism, in addition to the
gauge connection, a multiplet of scalar fields is introduced. These fields
parametrize the coset space between the gauge group and the stability
subgroup, play the role of Goldstone fields, and allow new identifications of
the vielbein and the spin connection in terms of the gauge fields and them
that, consequently, transform differently under the action of the elements of
the group. Thus, the presence of the Goldstone fields makes possible the
writing of the Einstein--Hilbert action in a way that is fully invariant under
the transformations of the Poincar\'{e} group (or AdS in the case of negative
cosmological constant), extending its local invariance under the Lorentz group
to the translation space. The same results can be found in the context of the
Poincar\'{e} and AdS supergravity theories.

On the other hand, first attempts of constructing non-abelian gauge theories
with higher-degree differential forms were carried out in ref.
\cite{Freedman:1980us} as generalizations of the Yang--Mills theory and
non-linear sigma models that are able to describe the coupling of gauge fields
to extended objects, such as strings or branes, in analogy to the standard
coupling of vector fields to point particles. Later studies shown that, when
gauging Lie algebras, the coupling of $p$-forms is compatible with spacetime
locality only for abelian gauge groups
\cite{Teitelboim:1985ya,Henneaux:1986ht}. The mathematical structure that
describe the gauge symmetry of these theories is a Lie algebra. However, the
presence of higher-degree forms as fundamental fields motivates to generalize
Lie algebras to structures that naturally include higher-degree forms in their
gauging. Lie algebras can be extended to free differential algebras (FDAs) by
means of their own topologial properties. They are extensions of the dual
formulation of Lie algebras in terms of Maurer--Cartan equations that are
introduced by including higher-degree differential forms as potentials, in
analogy to the Maurer--Cartan left-invariant one-forms, and demanding a
well-defined differential calculus in the corresponding manifold by means of
the integrability condition $\mathrm{d}^{2}=0$
\cite{DAuria:1980cmy,DAuria:1982uck}. Their mathematical consistency allows an
extension of the gauge principle that naturally includes higher-degree
differential forms in the field content a physical theory. This makes FDAs
particularly relevant in the construction of supergravity theories in
dimensionalities equal or higher than six, when the field contents require the
presence of three-forms in order to fulfill the supersymmetry requirement of
an equal number of degrees of freedom in the bosonic and fermionic sectors
\cite{castellani1991supergravity}. As it happens with Lie algebras, FDAs can
be alternatively formulated in terms of dual algebras. These algebras, known
as $L_{\infty}$ algebras, consist in graded vector spaces equipped multilinear
products that obey some specific rules of symmetry and antisymmetry, and the
so-called $L_{\infty}$ identities \cite{Chevalley:1948zz,Hohm:2017pnh}.

For a given symmetry, there are different procedures to construct an invariant
action principle. One of them is given by Chern--Simons theories, which are
topological theories that make use of Chern--Simons forms as Lagrangian
densities. Their gauge symmetry is described by Lie algebras and, although
they have been extensively studied in three dimensions for the cases of the
Poincar\'{e} and AdS groups due to the celebrated gravity quantization by
Edward Witten, they can be formulated in arbitrary odd-dimensionalities and
for any Lie group \cite{Witten:1988hc}. The reason for their
odd-dimensionality restriction lies in the topological origin of Chern--Simons
forms; they are locally-defined forms derived from a globally-defined
topological invariant density of one differential-degree higher, which is in
turn defined as proportional to a power of the two-form field strength, and
therefore, is always a differential form of even degree. For details on the
origin and application of Chern--Simons gravity and supergravity theories, see
refs.
\cite{Deser:1981wh,Achucarro:1986uwr,Chamseddine:1989nu,Chamseddine:1990gk,Banados:1996hi,Zanelli:2005sa,Izaurieta:2005vp,Mora:2006ka,Zanelli:2012px,Hassaine:2016amq}%
. The presence of higher-degree differential forms as fundamental fields in
the so-called higher gauge theories allows the existence of Chern--Simons
forms in both odd and even dimensionalities, since the building blocks of the
corresponding invariant densities are not two-forms anymore. This has been the
origin of different developments in the field of Chern--Simons theories with
$p$-forms. Examples of this kind can be found in refs.
\cite{Banados:1997qs,Antoniadis:2013jja,Antoniadis:2012ep,Konitopoulos:2014owa,Salgado:2017goq}%
, where differential forms in the adjoint representation of a Lie algebras are
included in the construction of several action principles. The symmetry
algebras of these theories can be interpreted as FDAs that carry no
representatives of a Chevalley--Eilenberg cohomology class in the construction
of the extended Maurer--Cartan equations or, equivalently, as $L_{\infty}$
algebras in which every multilinear product of three or more vectors vanishes.
For details on the formulation of $L_{\infty}$ algebras, their role on
classical gauge theories, and their relation with differential algebras, see
refs.
\cite{Hohm:2017pnh,Lada:1992wc,Lada:1994mn,Barnich:1997ij,Fiorenza:2013nha,Fiorenza:2016oki,Hohm:2017cey,Jurco:2018sby}%
. The inclusion of a cohomology class provides aditional structure to the
gauge algebra, making the symmetry richer, and also allows the construction of
new Chern--Simons action principles \cite{Salgado:2021eli,Salgado:2021sjo}.
However, the presence of the cohomology class changes the invariance
properties of the scalar inner product allowed by the algebra, making more
challenging the finding of an invariant tensor, and consequently, a gauge
invariant action principle. In this paper, we address the goal of extending
the SW-GN formalism as an alternative approach in the construction of action
principles invariant under the transformations of FDAs, in particular in the
cases in which the gauge algebra is a FDA constructed including one
representative of a non-trivial cohomology class into its Maurer--Cartan equations.

This article is structured as follows: In section 2 we review the basic
concepts from the literature on the SW-GN formalism in the context of Lie
gauge theories and present an introduction to the theory of FDAs, with
particular emphasis in the case known as FDA1, and its dual formulation. In
section 3 we extend the SW-GN formalism to the theory of FDAs, studying
separately the cases with and without non-trivial cohomology. In section 4 we
study two particular cases; we present non-trivial extensions of the
Poincar\'{e} and Maxwell bosonic Lie algebras, write down their dual
$L_{\infty}$ algebras and perform non-linear realizations that make possible
the construction of invariant action principles. In section 5, we derive the
$L_{\infty}$ algebra dual to the FDA of eleven-dimensional supergravity, and
formulate such theory as an off-shell gauge invariant theory of the mentioned
FDA. Finally, in section 6 we present the final comments about the research
and possible future developments. We include an appendix with some
calculations concerning the mentioned particular cases.

\section{Non-linear realizations and FDAs}

\subsection{The SW-GN formalism}

The SW-GN formalism consists in the use of non-linear realizations of Lie
groups to extend the gauge symmetry of a classical theory. With the purpose of
introducing these concepts, let us consider a Lie group $G$ with Lie algebra
$\mathfrak{g}$, and a subgroup $H\subset G$ with Lie algebra $\mathfrak{h}$.
Let $\left\{  V_{i}\right\}  _{i=1}^{\dim H}$ be a basis for $\mathfrak{h}$
and $\left\{  A_{l}\right\}  _{l=1}^{\dim G-\dim H}$ be the generators of the
remaining subspace, such that the set $\left\{  V_{i},A_{l}\right\}  $ is a
basis of $\mathfrak{g}$. We consider that the basis can be chosen such that
the generators $A_{l}$ form a representation of $H$, and therefore the Lie
products $\left[  V_{i},A_{l}\right]  $ are linear combinations of the vectors
$A_{l}$. An arbitrary group element $g_{0}$ can be decomposed in terms the
mentioned substructure as $g=e^{\xi^{l}A_{l}}h$, with $h\in H$, where the
zero-forms $\xi^{l}$ play the role of coordinates that parametrize the coset
space $G/H$. Since $g_{0}e^{\xi^{l}A_{l}}$ is also an element of $G$, it can
be decomposed as%
\begin{equation}
g_{0}e^{\xi^{l}A_{l}}=e^{\xi^{\prime l}A_{l}}h_{1}. \label{law}%
\end{equation}
The action of an arbitrary group element $g_{0}$ on $G/H$ is given by the
transformation law $\xi\rightarrow\xi^{\prime}$ in eq. (\ref{law}). Thus, eq.
(\ref{law}) allows to obtain the non-linear functions $\xi^{\prime}%
=\xi^{\prime}\left(  g_{0},\xi\right)  $ and $h_{1}=h_{1}\left(  g_{0}%
,\xi\right)  $. Let us now consider the case in which $g_{0}=h_{0}\in H$. In
this case, eq (\ref{law}) becomes%
\begin{equation}
e^{\xi^{\prime}}\equiv e^{\xi^{\prime l}A_{l}}=\left(  h_{0}e^{\xi^{l}A_{l}%
}h_{0}^{-1}\right)  h_{0}h_{1}^{-1}, \label{6}%
\end{equation}
and since the Lie product $\left[  V_{i},A_{l}\right]  $ is proportional to
$A_{l}$, one gets $h_{0}=h_{1}$ and the transformation law becomes linear:
\begin{equation}
e^{\xi^{\prime}}=h_{0}e^{\xi}h_{0}^{-1}. \label{5B}%
\end{equation}
On the other hand, if we consider that $g_{0}$ belongs to the coset space
$G/H$, eq. (\ref{law}) becomes
\begin{equation}
e^{\xi^{\prime l}A_{l}}=e^{\varepsilon^{l}A_{l}}e^{\xi^{l}A_{l}}h^{-1},
\label{5B2}%
\end{equation}
which is a non-linear transformation law for $\xi$. Moreover, by considering
that the transformation law $\xi\rightarrow\xi^{\prime}$ is described by the
infinitesimal variation $\delta$, eq. (\ref{law}) leads to
\begin{equation}
e^{-\xi^{l}A_{l}}\left(  g_{0}-1\right)  e^{\xi^{l}A_{l}}-e^{-\xi^{l}A_{l}%
}\delta e^{\xi^{l}A_{l}}=h_{1}-1. \label{5C}%
\end{equation}
Notice that, since $g_{0}-1$ is infinitesimal, $h_{1}-1$ is a vector of
$\mathfrak{h}$. Thus, the transformation law of $\xi$ can be obtained from the
requirement of components of the l.h.s. of eq. (\ref{5C}) along the generators
of the coset space to be vanishing.

Let us now introduce a one-form gauge connection $\mu$ evaluated on
$\mathfrak{g}$ and an action principle $S=S\left[  \mu\right]  $ which is
invariant under the stability subgroup $H$ but not under the transformations
along the generators of the coset space. Under the action of an arbitrary
group element $g_{0}=e^{-\varepsilon}$, the gauge connection transforms as%
\begin{equation}
\mu\longrightarrow\mu^{\prime}=e^{-\varepsilon}\mathrm{d}e^{\varepsilon
}+e^{-\varepsilon}\mu e^{\varepsilon}. \label{11}%
\end{equation}
We split $\mu$ into its contributions belonging in $\mathfrak{h}$ and the
coset space as $\mu=a+\rho$, with%
\begin{equation}
a=a^{l}A_{l},\text{ \ \ \ \ }\rho=\rho_{i}V_{i}.
\end{equation}
Moreover, we introduce a group element $z=e^{\xi}\equiv e^{\xi^{l}A_{l}}$ and
define the non-linear gauge connection%
\begin{equation}
\mu^{\xi}=\mu^{\xi}\left(  \mu,\xi\right)  =z^{-1}\mathrm{d}z+z^{-1}\mu z.
\label{12}%
\end{equation}
The functional form of $\mu^{\xi}$ is given by a large gauge tranformation of
$\mu$ with group element $z^{-1}$, that non-linearly depends on the zero-forms
$\xi^{l}$ and their derivatives. However, $\mu^{\xi}$ is interpreted as the
fundamental field of a gauge theory and therefore, both $\mu$ and $\xi$ change
under the action of the gauge group. As before, we split the contributions to
the non-linear connection as $\mu^{\xi}=v+p$ with%
\begin{equation}
p=p^{l}\left(  \xi,\mathrm{d}\xi\right)  A_{l},\text{ \ \ \ \ }v=v^{i}\left(
\xi,\mathrm{d}\xi\right)  V_{i}.
\end{equation}
It is possible to prove that, under the change $\xi\rightarrow\xi^{\prime}$,
the transformation laws for $p$ and $v$ are given by%
\begin{align}
p  &  \longrightarrow p^{\prime}=h_{1}ph_{1}^{-1},\label{p}\\
v  &  \longrightarrow v^{\prime}=h_{1}vh_{1}^{-1}+h_{1}\mathrm{d}h_{1}^{-1},
\label{v}%
\end{align}
i.e., when acting with $g_{0}\in G/H$, the non-linear one-forms $p$ and $v$
transform as a tensor and as a connection respectively. These transformations
are linear but the group element is now a function of the parameters, i.e.,
$h_{1}=h_{1}\left(  \varepsilon,\xi\right)  $. Eqs. (\ref{p}) and (\ref{v})
show that, any quantity defined in terms of $\mu$ that is invariant under the
stability subgroup $H$, will become invariant under $G$ if it is written in
terms of $\mu^{\xi}$. Notice that the gauge transformation law of the scalar
multiplet $\xi\,$\ under the transformations of the stability subalgebra and
the coset space are given by eqs. (\ref{5B}) and (\ref{5C}) respectively. By
considering that the gauge parameter $\varepsilon$ is infinitesimal, it is
possible to write down these laws in terms of the Lie product. In fact, by
consider an infinitesimal gauge parameter $\varepsilon=\varepsilon^{i}V_{i}%
\in\mathfrak{h}$, eq. (\ref{5B}) is reduced to%
\begin{equation}
\delta\xi=\left[  \xi,\varepsilon\right]  . \label{5Binf}%
\end{equation}
On the other hand, by considering an infinitesimal gauge parameter
$\varepsilon=\varepsilon^{l}A_{l}\in\mathfrak{g}/\mathfrak{h}$, the
transformation law (\ref{5C}) can be written in an equivalent way that makes
no explicit use of an exponential map. To this end, let us recall the
following identities:%
\begin{align}
e^{X}Ye^{-X}  &  =\sum_{n=0}^{\infty}\frac{1}{n!}\left[  X,\overset{n}{\ldots
},\left[  X,Y\right]  \right]  ,\label{id1}\\
e^{X}\delta e^{-X}  &  =-\sum_{n=1}^{\infty}\frac{1}{n!}\left[
X,\overset{n-1}{\ldots},\left[  X,Y\right]  \right]  . \label{id2}%
\end{align}
Here, $X$ and $Y~$\ are two arbitrary vectors of $\mathfrak{g}$ and $\delta$
is an arbitrary variation. By applying the identities (\ref{id1}) and
(\ref{id2}) into the transformation law (\ref{5C}), one finds that it can be
written in a more convenient way
\begin{equation}
\left.  \sum_{n=0}^{\infty}\left(  \frac{\left(  -1\right)  ^{n}}{n!}\left[
\xi,\overset{n}{\cdots},\left[  \xi,\varepsilon\right]  \right]
+\frac{\left(  -1\right)  ^{n}}{\left(  n+1\right)  !}\left[  \xi
,\overset{n}{\cdots},\left[  \xi,\delta\xi\right]  \right]  \right)
\right\vert _{\mathfrak{g}/\mathfrak{h}}=0, \label{5Cinf}%
\end{equation}
As it happens in eq. (\ref{5C}), eq. (\ref{5Cinf}) states that the result of
the sum must be a vector of $\mathfrak{h}$. This means that its component
along $\mathfrak{g/h}$ vanishes. It is important to point out that the
connection $\mu^{\xi}$ transforms according to (\ref{p}) and (\ref{v}), if and
only if the variation of the multiplet $\xi$ satisfies eqs. (\ref{5Binf}) and
(\ref{5Cinf}). In other words, the vanishing of the components along
$\mathfrak{g/h}$ in eq. (\ref{5Cinf}) ensures that $\mu^{\xi}$ transforms as a
connection of $\mathfrak{h}$, even if the parameter of the transformation lies
in the coset space. Moreover, eq. (\ref{5Cinf}) allows to explicitly obtain
the components of $\delta\xi$.

As we have seem the transformation law of the multiplet can be convenietly
written in terms of the algebraic product. The same happens with the large
gauge transformation of the gauge connection. In fact, by applying the
identities (\ref{id1}) and (\ref{id2}) into the general expression
(\ref{11})\thinspace we find that the gauge transformed is given by%
\begin{align}
\mu^{\prime}  &  =\sum_{n=0}^{\infty}\frac{\left(  -1\right)  ^{n}}{n!}\left[
\varepsilon,\overset{n}{\ldots},\left[  \varepsilon,\mu\right]  \right]
-\sum_{n=1}^{\infty}\frac{\left(  -1\right)  ^{n}}{n!}\left[  \varepsilon
,\overset{n-1}{\ldots},\left[  \varepsilon,\mathrm{d}\varepsilon\right]
\right] \nonumber\\
&  =\mu+\sum_{n=1}^{\infty}\frac{\left(  -1\right)  ^{n+1}}{n!}\left[
\varepsilon,\overset{n-1}{\ldots},\left[  \varepsilon,\nabla\varepsilon
\right]  \right]  \label{large1}%
\end{align}
Eq. (\ref{large1}) can be used as an alternative definition of large gauge
transformation that only makes use of the algebraic product and does not
require a notion of exponentiation.

As an example, let us consider that $G$ is the four-dimensional Poincar\'{e}
group spanned by the translation generators $P_{a}$ and the Lorentz rotation
generators $J_{ab}$ with $a,b=0,\ldots3$. Let $H$ be the Lorentz subgroup. In
this case, we denote the components of the linear and non-linear connections
as%
\begin{align}
\mu &  =e^{a}P_{a}+\frac{1}{2}\omega^{ab}J_{ab},\\
\mu^{\xi}  &  =V^{a}P_{a}+\frac{1}{2}W^{ab}J_{ab}.
\end{align}
According to eq. (\ref{large1}),\ the non-linear connection $\mu^{\xi}$ is
given by%
\begin{equation}
\mu^{\xi}=V^{a}P_{a}+\frac{1}{2}W^{ab}J_{ab},
\end{equation}
with%
\begin{align}
V^{a}  &  =e^{a}+\mathrm{D}_{\omega}\xi^{a},\\
W^{ab}  &  =\omega^{ab},
\end{align}
where $\mathrm{D}_{\omega}$ is the covariant derivative defined with the
Lorentz connection $\omega^{ab}$. Eqs. (\ref{5Binf}) and (\ref{5Cinf}) allow
to obtain the transformation laws of the scalar multiplet under the
transformations of the stability subgroup and the coset space respectively. In
this case, by considering a zero-form gauge parameter%
\begin{equation}
\varepsilon=\varepsilon^{a}P_{a}+\frac{1}{2}\varepsilon^{ab}J_{ab},
\end{equation}
one finds that the components of the scalar multiplet transform as
\begin{equation}
\delta\xi^{a}=-\varepsilon^{a}-\varepsilon_{\text{ \ }c}^{a}\xi^{c}.
\end{equation}
Moreover, according to eq. (\ref{large1}), the components of $\mu$ transform
as $\delta e^{a}=\mathrm{D}_{\omega}\varepsilon^{a}-\varepsilon_{\text{ \ }%
c}^{a}e^{c}$ and $\delta\omega^{ab}=\mathrm{D}_{\omega}\varepsilon^{ab}$. As a
consecuence, $V^{a}$ transforms as $\delta V^{a}=-\varepsilon_{\text{ \ }%
c}^{a}V^{c}$, i.e., it changes as a vector under Lorentz rotations while
remains invariant under translations. Thus, by replacing $\mu$ by $\mu^{\xi}$,
every Lorentz invariant form will become invariant under the action of the
Poincar\'{e} group. This allows to formulate four-dimensional general
relativity as a gauge theory of the Poincar\'{e} group by identifying the
vielbein field one form as $V^{a}$ instead of $e^{a}$. The zero-forms $\xi
^{a}$ play the role of Goldstone fields and have no dynamics in the resulting
equations of motion \cite{Salgado:2001bn,Salgado:2003rf}.

\subsection{FDAs and $L_{\infty}$ algebras}

Let us now consider a brief introduction to the theory of free differential
algebras and its relation with $L_{\infty}$ algebras, in particular for the
case known as FDA1. FDAs are defined as extensions of the dual formulations of
Lie algebras. That formulation allows to define a Lie algebra $\mathfrak{g}$
in terms of a set of differential equations (or Maurer--Cartan equations) for
one-forms defined in the corresponding soft manifold $M\,$\ where, for
consistency, the differential operator $\mathrm{d}$ must satisfy the
integrability condition $\mathrm{d}^{2}=0\,$. In analogy, a FDA is defined in
terms of differential equations for a set of forms of different degrees
$\left\{  \Theta^{A\left(  p\right)  }\right\}  _{p=1}^{N}$, where each
$\Theta^{A\left(  p\right)  }$ is a $p$-form, $N$ is the degree of the highest
form considered, which is set as the dimensionality of $M$, and each index
$A\left(  p\right)  $ takes values in a different domain. Thus, two indices
$A\left(  p\right)  $ and $A\left(  q\right)  $ are in general not
contractible unless $p=q$. Since the set $\left\{  \Theta^{A\left(  p\right)
}\right\}  $ is a basis, it becomes possible to write the exterior derivative
of each form in terms of the others:
\begin{equation}
\mathrm{d}\Theta^{A\left(  p\right)  }+\sum_{n=1}^{N}\frac{1}{n}%
C_{B_{1}\left(  p_{1}\right)  \cdots B_{n}\left(  p_{n}\right)  }^{A\left(
p\right)  }\Theta^{B_{1}\left(  p_{1}\right)  }\wedge\cdots\wedge\Theta
^{B_{n}\left(  p_{n}\right)  }=0. \label{fda6}%
\end{equation}
Eq. (\ref{fda6}) is known as generalized Maurer--Cartan equation and it
defines a general FDA as long as the structure constants $C_{B_{1}\left(
p_{1}\right)  \cdots B_{n}\left(  p_{n}\right)  }^{A\left(  p\right)  }$
satisfy some consistency conditions. Notice that, in contrast with the Lie
algebra structure constants, the second term in the l.h.s. of eq. (\ref{fda6})
allows the presence of differential forms of odd and even degrees. As a
consequence, the lower indices of the structure constants $C_{B_{1}\left(
p_{1}\right)  \cdots B_{n}\left(  p_{n}\right)  }^{A\left(  p\right)  }$ can
be both symmetric or antisymmetric. In analogy to the Lie algebra case, the
integrability condition leads to a set of generalized Jacobi identities%
\begin{equation}
\sum_{m,n=1}^{N}\frac{1}{m}C_{B_{1}\left(  p_{1}\right)  \cdots B_{n}\left(
p_{n}\right)  }^{A\left(  p\right)  }C_{C_{1}\left(  q_{1}\right)  \cdots
C_{m}\left(  q_{m}\right)  }^{B_{1}\left(  p_{1}\right)  }\Theta^{C_{1}\left(
q_{1}\right)  }\wedge\cdots\wedge\Theta^{C_{m}\left(  q_{m}\right)  }%
\wedge\Theta^{B_{2}\left(  p_{2}\right)  }\wedge\cdots\wedge\Theta
^{B_{n}\left(  p_{n}\right)  }=0. \label{Jacobi}%
\end{equation}
Eq. (\ref{Jacobi}) can be written in terms of the structure constants by
removing the basis of differential forms and including the corresponding signs
of symmetry or antisymmetry. Under a general permutation $\sigma$ of $n$
elements, their symmetry rule is given by%
\begin{equation}
C_{B_{1}\left(  p_{1}\right)  \cdots B_{n}\left(  p_{n}\right)  }^{A\left(
p\right)  }=\epsilon\left(  \sigma,p\right)  C_{B_{\sigma\left(  1\right)
}\left(  p_{\sigma\left(  1\right)  }\right)  \cdots B_{\sigma\left(
n\right)  }\left(  p_{\sigma\left(  n\right)  }\right)  }^{A\left(  p\right)
},
\end{equation}
where $\epsilon\left(  \sigma,p\right)  $ is the Koszul sign depending on
$\sigma$ and the set of labels $p=\left\{  p_{1},...,p_{n}\right\}  $. The
Koszul sign is in general defined according the following rule: Let $\left\{
A_{i}\right\}  _{i=1}^{n}$ be a set of differential forms of differential
degrees $a_{i}$, the Koszul sign is the factor that appears in the permutation%
\begin{equation}
A_{i_{1}}\wedge\cdots\wedge A_{i_{n}}=\epsilon\left(  \sigma,a\right)
A_{i_{\sigma\left(  1\right)  }}\wedge\cdots\wedge A_{i_{\sigma\left(
n\right)  }},
\end{equation}
which depends on the permutation $\sigma$ and the degrees of the forms $a_{i}$
\cite{Hohm:2017pnh,Hohm:2017cey}.

Let us now introduce a dual bracket, in analogy to the Lie bracket of a Lie
algebra. Let $\bar{X}$ be a graded vector space and $\left\{  t_{A\left(
p\right)  }\right\}  _{p=1}^{N}$ be a basis for $\bar{X}$ with $\deg_{\bar{X}%
}t_{A\left(  p\right)  }=p$, where the indices $A\left(  p\right)  $ satisfy
the same structure introduced above for a general FDA. We define a set of
$n$-linear products ($n\geq1$) acting on $\bar{X}$%
\begin{equation}
\left[  t_{A_{1}\left(  p_{1}\right)  },\ldots,t_{A_{n}\left(  p_{n}\right)
}\right]  _{n}\in\bar{X}. \label{Tprod}%
\end{equation}
Since each product $\left[  t_{A_{1}\left(  p_{1}\right)  },\ldots
,t_{A_{n}\left(  p_{n}\right)  }\right]  _{n}$ is also a vector in $\bar{X}$,
they can be written in terms of linear combiations of $t_{A\left(  p\right)
}$. We define their components as proportional to the structure constants of a
FDA, as follows:%
\begin{equation}
\left[  t_{A_{1}\left(  p_{1}\right)  },\ldots,t_{A_{n}\left(  p_{n}\right)
}\right]  _{n}^{A\left(  p\right)  }=\left(  n-1\right)  !C_{A_{1}\left(
p_{1}\right)  \cdots A_{n}\left(  p_{n}\right)  }^{A\left(  p\right)  },
\label{T}%
\end{equation}
and therefore, verifying the same graded-symmetry relation%
\begin{equation}
\left[  t_{A_{\sigma\left(  1\right)  }\left(  p_{\sigma\left(  1\right)
}\right)  },\ldots,t_{A_{\sigma\left(  n\right)  }\left(  p_{\sigma\left(
n\right)  }\right)  }\right]  _{n}=\epsilon\left(  \sigma,t\right)  \left[
t_{A_{1}\left(  p_{1}\right)  },\ldots,t_{A_{n}\left(  p_{n}\right)  }\right]
_{n}. \label{symmb}%
\end{equation}
Notice that, for $n=2$ and by setting $p_{1}=p_{2}=1$, one recovers the
antisymmetric bilinear product of a Lie algebra.

Eq. (\ref{T}) allows to write the generalized Jacobi identity from eq.
(\ref{Jacobi}) in terms of the multilinear products. We can now remove the
dependence on the differential forms and add the corresponding sign depending
on the symmetry or antisymmetry of their permutation inside the sum. Thus, we
rename the indices and perform the sum over $m!\left(  n-1\right)  !$
equivalent elements. This reduces eq. (\ref{Jacobi}) to%
\begin{equation}
\sum_{m+n=l-1}^{N}\sum_{\sigma\in U\left(  l\right)  }\epsilon\left(
\sigma,t\right)  \left[  \left[  t_{B_{\sigma\left(  1\right)  }\left(
q_{\sigma\left(  1\right)  }\right)  },\ldots,t_{B_{\sigma\left(  m\right)
}\left(  q_{\sigma\left(  m\right)  }\right)  }\right]  _{m},t_{B_{\sigma
\left(  m+1\right)  }\left(  q_{\sigma\left(  m+1\right)  }\right)  }%
\ldots,t_{B_{\sigma\left(  l\right)  }\left(  p_{\sigma\left(  l\right)
}\right)  }\right]  _{n}=0, \label{jacb}%
\end{equation}
where now the first sum runs over unsuffles, i.e., over permutations
satisfying $\sigma\left(  1\right)  <\cdots<\sigma\left(  m\right)  $ and
$\sigma\left(  m+1\right)  <\cdots<\sigma\left(  l\right)  $
\cite{Hohm:2017pnh,Hohm:2017cey}. Eq. (\ref{jacb}) is known as $L_{\infty}$
identity in the $b$-picture and it is equivalent to the Jacobi identity in
dual formulation. Thus, any FDA can be written in dual formulation as an
$L_{\infty}$ algebra. Notice that eq. (\ref{jacb}) reproduces the standard
Jacobi identity for a Lie algebra in the case in which $\bar{X}$ is given by a
single subspace of degree $1$. The dual relation between FDAs and $L_{\infty}$
algebras has been pointed out in refs.
\cite{Fiorenza:2013nha,Fiorenza:2016oki}. For an extensive review, see also
ref. \cite{Fiorenza:2019ckz}

The FDA1 algebra is a particular FDA, in which the only non-vanishing
potentials are a one-form, denoted by $\mu^{A}$, and a $p$-form, denoted by
$\mu^{i}$, and therefore described by the following Maurer--Cartan equations
\cite{Castellani:1995gz,Castellani:2006jg,Castellani:2005vt}%
\begin{align}
\mathrm{d}\mu^{A}+\frac{1}{2}C_{BC}^{A}\mu^{B}\mu^{C}  &  =R^{A}%
=0,\label{mu}\\
\mathrm{d}\mu^{i}+C_{Aj}^{i}\mu^{A}\mu^{j}+\frac{1}{\left(  p+1\right)
!}C_{A_{1}\cdots A_{p+1}}^{i}\mu^{A_{1}}\cdots\mu^{A_{p+1}}  &  =R^{i}=0.
\label{B}%
\end{align}
Notice that structure constants of the type $C_{A}^{i}$ are not included,
although they are in general allowed for $p=2$. Algebras that share this
feature are called minimal. Moreover, we have renamed the indices as $A\left(
1\right)  \equiv A$ and $A\left(  p\right)  \equiv i$. The Jacobi identities
are reduced to three independent conditions for the FDA1 structure constants.
The first one is the standard Jacobi identity of a Lie algebra for $C_{BC}%
^{A}$. This shows that FDA1 has the a Lie subalgebra, defined by eq.
(\ref{mu}). The second Jacobi identity is the statement that the structure
constants $C_{Aj}^{i}$ form a representation of that Lie subalgebra. The third
Jacobi identity is equivalent to the requirement of the last term in the l.h.s
of eq. (\ref{B}) to be closed under the covariant derivative defined with the
mentioned representation, i.e., to be a $\left(  p+1\right)  $-cocycle of the
Lie subalgebra \cite{Castellani:2013mka}. These identities allows to write eq.
(\ref{B}) as%
\begin{equation}
\left(  \nabla\mu\right)  ^{i}+\Omega^{i}=0, \label{Bv2}%
\end{equation}
where the $\left(  p+1\right)  $-form%
\begin{equation}
\Omega^{i}=\frac{1}{\left(  p+1\right)  !}C_{A_{1}\cdots A_{p+1}}^{i}%
\mu^{A_{1}}\cdots\mu^{A_{p+1}},
\end{equation}
satisfies $\nabla\Omega^{i}=0$. The requirement of covariant closure
classifies the cocycles of a Lie algebra in Chevalley--Eilenberg cohomology
classes, consisting of forms that are covariatly closed but not covariantly
exact. Thus, two cocycles belong to the same class if they differ in a
covariatly closed form. Consequently, two FDAs carrying cocycles belonging to
the same cohomology class are considered equivalent, since one can be obtained
from the other one through a redefinition of the Maurer--Cartan potentials.

The gauging of FDA1 implies to consider non-vanishing curvature $R=\left(
R^{A},R^{i}\right)  $, whose components are the two-form and $\left(
p+1\right)  $-form defined in eqs. (\ref{mu}) and (\ref{B}) respectively.
Moreover, in order to introduce gauge transformations we define a composite
gauge parameter $\varepsilon=\left(  \varepsilon^{A},\varepsilon^{i}\right)
$, where $\varepsilon^{A}$ is a zero-form and $\varepsilon^{i}$ is a
$(p-1)$-form. The infinitesimal gauge variations of the components of $\mu$
are given by%
\begin{align}
\delta\mu^{A}  &  =\mathrm{d}\varepsilon^{A}+C_{BC}^{A}\mu^{B}\varepsilon
^{C},\label{var1}\\
\delta\mu^{i}  &  =\mathrm{d}\varepsilon^{i}+C_{Aj}^{i}\mu^{A}\varepsilon
^{j}-C_{Aj}^{i}\varepsilon^{A}\mu^{j}-\frac{1}{p!}C_{AA_{1}\cdots A_{p}}%
^{i}\varepsilon^{A}\mu^{A_{1}}\cdots\mu^{A_{p}}. \label{var2}%
\end{align}

With the purpose of writing down the dual $L_{\infty}$ algebra, we introduce
the vectors $t_{A}$ generating the Lie subalgebra and $t_{i}$ as duals to
$\mu^{i}$, lying in the following subspaces%
\begin{equation}
t_{A}\in\bar{X}_{1},\text{ \ \ \ }t_{i}\in\bar{X}_{p}.
\end{equation}
According to eq. (\ref{T}), the dual $L_{\infty}$ algebra is defined as the
graded vector space $\bar{X}=\bar{X}_{1}\oplus\bar{X}_{p}$ equipped with the
following bilinear and $\left(  p+1\right)  $-linear products:%
\begin{align}
\left[  t_{B},t_{C}\right]  _{2}  &  =-\left[  t_{C},t_{B}\right]  _{2}%
=C_{BC}^{A}t_{A},\label{fda1A}\\
\left[  t_{A},t_{j}\right]  _{2}  &  =\left(  -1\right)  ^{p}\left[
t_{j},t_{A}\right]  _{2}=C_{Aj}^{i}t_{i},\label{fda1B}\\
\left[  t_{A_{1}},\ldots,t_{A_{p+1}}\right]  _{p+1}  &  =p!C_{A_{1}\cdots
A_{p+1}}^{i}t_{i}. \label{fda1C}%
\end{align}
We denote this dual algebra as $L_{\infty}^{\mathrm{FDA1}}$. The writting of
FDA1 in terms of its dual algebra turns out to be useful in further calculations.

\section{The SW-GN formalism in FDA theory}

The SW-GN formalism consists in an extension of the gauge symmetry of a
classical gauge theory from a stability algebra $\mathfrak{h}$ to a
higher-dimensional one $\mathfrak{g}$, which contains $\mathfrak{h}$ as a
subalgebra. This is carried out by considering a gauge connection $\mu$
evaluated in the full algebra and an action principle $I\left[  \mu\right]  $
presenting invariance under the transformations of the stability algebra. The
invariance of the action is then extended by introducing a scalar multiplet
$\xi$ evaluated in the coset space $\mathfrak{g/h}$, and defining a non-linear
gauge field $\mu^{\xi}$ (see eq. (\ref{12})). The non-linear gauge field is
mathematically introduced by means of a gauge transformation of the original
one, where $\xi$ is the parameter of the transformation. Then, the action is
written as a functional of $\mu^{\xi}$ in direct replacement of $\mu$. The
symmetry of the action is extended by fixing the transformation law of $\xi$
in a convenient way, specifically given by (\ref{5C}). As it was shown in the
example of section 2.1, the SW-GN formalism can be used to extend the Lorentz
invariance of the four-dimensional Einstein--Hilbert action. In that case, the
stability algbera is identified as the Lorentz algebra, while the full algebra
is the Poincar\'{e} algebra. Thus, general relativity can be forumated as a
genuine gauge theory of the Poincar\'{e} group. In this section, we aim to
extend this formalism to the case of FDA1. However, it must be noticed that
this formalism requires a definition of large gauge transformation in order to
define the non-linear gauge field. Such definition makes use an exponentiation
mapping, which in general cannot be defined for FDAs. This is a consequence of
the presence of higher-degree forms in the gauge parameters. In order to
define large gauge transformations for FDA1, we present the following construction.

Let us first consider the gauging of FDA1 in terms of the dual vectors $t_{A}$
and $t_{i}$. For this purpose, let us introduce the contraction operators
$\imath_{t_{A}}$ and $\imath_{t_{i}}$ in terms of their action on the FDA1
potentials \cite{Castellani:1995gz,Castellani:2006jg,Castellani:2005vt}:%
\begin{align}
\imath_{A}\mu^{B}  &  =\delta_{A}^{B},\text{ \ \ \ \ }\imath_{A}\mu
^{i}=0,\label{ii1}\\
\imath_{i}\mu^{B}  &  =0,\text{ \ \ \ \ }\imath_{i}\mu^{j}=\delta_{i}^{j},
\label{ii2}%
\end{align}
which, by definition satisfy the Leibniz rule. Since $\mu^{A}$ and $\mu^{i}$
are a basis of differential forms, every other form can be written as a sum of
polynomials in $\mu^{A}$ and $\mu^{i}$. The operations $\imath_{A}$ and
$\imath_{i}$ act on the algebra of polynomials spanned by $\left\{  \mu
^{A},\mu^{i}\right\}  $ by decreasing the differential degree by $1$ and $p$
respectively. Moreover, since the FDA1 gauge parameter $\varepsilon=\left(
\varepsilon^{A},\varepsilon^{i}\right)  $ is composed by a zero-form and a
$\left(  p-1\right)  $-form, it becomes possible to introduce a general
contraction with $\varepsilon$ as%
\begin{equation}
\imath_{\varepsilon}=\varepsilon^{A}\imath_{A}+\varepsilon^{i}\imath_{i},
\label{sum}%
\end{equation}
such that the resulting operator reduces the differential degree of the form
on which it acts by $1$. Furthermore, by considering the FDA1 contraction
operators $\imath_{A}$ and $\imath_{i}$, it is possible to define generalized
Lie derivatives for FDA1 as%
\begin{equation}
\ell_{A}=\mathrm{d}\imath_{A}+\imath_{A}\mathrm{d},\text{ \ \ \ \ }\ell
_{i}=\mathrm{d}\imath_{i}+\imath_{i}\mathrm{d}.
\end{equation}
The Lie derivatives $\ell_{A}$ along the standard directions $t_{A}$ leave
invariant the differential degree of the forms on which they act, just as it
happens in the study of Lie algebras. In contrast, the derivatives $\ell_{i}$
along the extended directions $t_{i}$ decrease the differential degree of the
form on which it acts by $p-1$. We identify these operators with the basis
vectors of $L_{\infty}^{\mathrm{FDA1}}$, i.e.,%
\begin{equation}
t_{A}\equiv\ell_{A},\text{ \ \ \ \ }t_{i}\equiv\ell_{i} \label{identify}%
\end{equation}
Thus, the $L_{\infty}$ algebra from eqs. (\ref{fda1A})-(\ref{fda1C}) is
defined as the vector space of Lie derivatives endowed with a set of bilinear
and $\left(  p+1\right)  $-linear mappings between these operators.
Furthermore, this identification allows us to write the gauge field as a
single linear combination:
\begin{equation}
\mu=\mu^{A}t_{A}+\mu^{i}t_{i}=\mu^{I}t_{I}, \label{one}%
\end{equation}
where we introduce the composite index $I$, taking values in both algebraic
sectors, the standard one spanned by $t_{A}$ an the extended one spanned by
$t_{i}$. It is importat to point out that, since $t_{i}$ carries differential
degree $\left(  1-p\right)  $, the sum in equation (\ref{one}) remains a
one-form. In the same way, we write the gauge curvature associated to $\mu$
and the infinitesimal gauge parameters as
\begin{align}
R  &  =R^{A}t_{A}+R^{i}t_{i},\label{Rfull}\\
\varepsilon &  =\varepsilon^{A}t_{A}+\varepsilon^{i}t_{i}. \label{efull}%
\end{align}
As it happens with the gauge field, the extended components $R^{i}$ and
$\varepsilon^{i}$ are differential forms of higher degree. However, their
contraction with the operator $t_{i}$ makes possible their sum with the
standard components $R^{A}t_{A}$ and $\varepsilon^{A}t_{A}$. The use of this
basis allows to write down the gauge curvatures and variations in a index free
way, i.e., independently of the chosen basis. In fact, we can now make use of
this notation to write the bilinear and $\left(  p+1\right)  $-linear products
$\left[  \mu,\mu\right]  _{2}$ and $\left[  \mu^{p+1}\right]  _{p+1}$ in terms
of their components in both algebraic sectors, as follows%
\begin{align}
\left[  \mu,\mu\right]  _{2}  &  =\left[  \mu,\mu\right]  _{2}^{A}%
t_{A}+\left[  \mu,\mu\right]  _{2}^{i}t_{i},\label{mu2a}\\
\left[  \mu^{p+1}\right]  _{p+1}  &  =\left[  \mu^{p+1}\right]  _{p+1}%
^{A}t_{A}+\left[  \mu^{p+1}\right]  _{p+1}^{i}t_{i}, \label{mu2b}%
\end{align}
where the contributions corresponding to the standard and extended sectors are
obtained by directly plugging in the relations of the $L_{\infty}$ algebra
(\ref{fda1A})-(\ref{fda1C}):
\begin{align}
\left[  \mu,\mu\right]  _{2}^{A}  &  =C_{BC}^{A}\mu^{B}\mu^{C},\text{
\ \ \ \ \ \ \ }\left[  \mu,\mu\right]  _{2}^{i}=2C_{Aj}^{i}\mu^{A}\mu
^{j},\label{mu2c}\\
\left[  \mu^{p+1}\right]  _{p+1}^{A}  &  =0,\text{ \ \ \ \ \ \ \ }\left[
\mu^{p+1}\right]  _{p+1}^{i}=p!C_{A_{1}\cdots A_{p+1}}^{i}\mu^{A_{1}}\cdots
\mu^{A_{p+1}}. \label{mu2d}%
\end{align}
Similarly, the bilinear and multilinear products $\left[  \varepsilon
,\mu\right]  _{2}$ and $\left[  \varepsilon,\mu^{p}\right]  _{p+1}$ carry the
following components%
\begin{align}
\left[  \varepsilon,\mu\right]  _{2}^{A}  &  =C_{BC}^{A}\varepsilon^{B}\mu
^{C},\text{ \ \ \ \ \ \ \ }\left[  \varepsilon,\mu\right]  _{2}^{i}=C_{Aj}%
^{i}\left(  \varepsilon^{A}\mu^{j}-\mu^{A}\varepsilon^{j}\right)
,\label{mue1}\\
\left[  \varepsilon,\mu^{p}\right]  _{p+1}^{A}  &  =0,\text{ \ \ \ \ \ \ \ }%
\left[  \varepsilon,\mu^{p}\right]  _{p+1}^{i}=p!C_{A_{1}\cdots A_{p+1}}%
^{i}\varepsilon^{A_{1}}\mu^{A_{2}}\cdots\mu^{A_{p+1}}t_{i}. \label{mue3}%
\end{align}
By plugging in eqs. (\ref{mu2a})-(\ref{mue3}) into eqs. (\ref{mu}), (\ref{B}),
(\ref{var1}) and (\ref{var2}), it is direct to see that this notation allows
to write the Maurer--Cartan equations and infinitesimal gauge transformations
in a more compact way that only makes use of the exterior derivative operator
and the $L_{\infty}^{\mathrm{FDA1}}$ products, and that does not depend on the
chosen basis:%
\begin{align}
R  &  =\mathrm{d}\mu+\frac{1}{2}\left[  \mu,\mu\right]  _{2}+\frac
{1}{p!\left(  p+1\right)  !}\left[  \mu^{p+1}\right]  _{p+1}=0, \label{Ralone}%
\\
\delta\mu &  =\mathrm{d}\varepsilon-\left[  \varepsilon,\mu\right]  _{2}%
-\frac{1}{p!p!}\left[  \varepsilon,\mu^{p}\right]  _{p+1}.
\label{deltamualone}%
\end{align}

It is important to point out that eq. (\ref{identify}) is not the only
possible identification of dual vectors. In fact, in refs.
\cite{Castellani:1995gz,Castellani:2006jg,Castellani:2005vt}, the Lie
derivavites along the directions $t_{A}$ and $\mu^{A_{1}}\cdots\mu^{A_{p-1}%
}t_{i}$ are identified as basis vectors generating an alternative dual
algebra. These algebras and $L_{\infty}$ algebras are non-associative and can
be considered as dual to FDA1, since they carry the information about the
structure constants and the cocycle. Thus, in contrast with Lie algebras, FDAs
present more than a single dual formulation in terms of vector spaces endowed
with linear products. The main difference between these two dual formulations
lies in the fact that $L_{\infty}$ algebras are equipped with multilinear
products, while the products of the mentioned alternative dual algebras are
always bilinear. For details on the non-associativity of these bilinear dual
algebras, see ref. \cite{Castellani:2013mka}.

\subsection{Trivial cohomology}

Let us now study the existence of large gauge transformations in the context
of FDA gauge theories. Let us recall that, in Lie group theory, the large
gauge transformations of the connection is given by eq. (\ref{11}), while the
corresponding gauge curvature transforms as%
\begin{equation}
R\rightarrow R^{\prime}=e^{-\varepsilon}Re^{\varepsilon}.
\end{equation}
Here $\varepsilon=\varepsilon^{A}t_{A}$ is a non-infinitesimal parameter and
$\left\{  t_{A}\right\}  $ are the generators of the Lie algebra in a chosen
basis. This transformation law can be immediately extended to the case of a
FDA1 that carries a $p$-form but no cocycle in the generalized Maurer--Cartan
equation. To see this, let us consider eqs. (\ref{mu}) and (\ref{B}) with
$C_{A_{1}\cdots A_{p+1}}^{i}=0$. The gauge parameter can be written in terms
of a composite index as in eq. (\ref{efull}). In this case, the gauge
transformations, both standard and extended, can be written as%
\begin{equation}
\delta\mu=\mathrm{d}\varepsilon+\left[  \mu,\varepsilon\right]  ,
\label{deltamufull}%
\end{equation}
or, in components as follows:%
\begin{align}
\delta\mu^{A}  &  =\mathrm{d}\varepsilon^{A}+C_{BC}^{A}\mu^{B}\varepsilon
^{C},\label{delta1}\\
\delta\mu^{i}  &  =\mathrm{d}\varepsilon^{i}+C_{Aj}^{i}\left(  \mu
^{A}\varepsilon^{j}-\varepsilon^{A}\mu^{j}\right)  . \label{delta2}%
\end{align}
Since the contraction $\varepsilon^{i}t_{i}$ carries no differential degree in
eq. (\ref{deltamufull}), it becomes possible to define an exponential of the
type $z=\exp\left(  \varepsilon\right)  $ such that, a finite gauge
transformation is given by eq. (\ref{12}). In other words, the gauging of this
FDA can be performed by considering a shifting in the gauge fields and
parameters by means of the new vectors $t_{i}$ on which the higher-degree
forms can be evaluated. In this case, the dual $L_{\infty}$ algebra given by
eqs. (\ref{fda1A})-(\ref{fda1C}) is reduced to
\begin{align}
\left[  t_{A},t_{B}\right]  _{2}  &  =C_{AB}^{C}t_{C},\\
\left[  t_{A},t_{i}\right]  _{2}  &  =C_{Ai}^{j}t_{j}.
\end{align}
Notice that the product $\left[  t_{A},t_{i}\right]  _{2}$ is symmetric or
antisymmetric for even and odd values of $p$ respectively. By directly
performing two sucesive gauge transformations, it is direct to verify that, in
the extended algebraic sector, they compose as%
\begin{gather}
\delta_{\varepsilon_{2}}\delta_{\varepsilon_{1}}-\delta_{\varepsilon_{1}%
}\delta_{\varepsilon_{2}}=\delta_{\varepsilon_{3}},\\
\varepsilon_{3}^{i}=C_{Aj}^{i}\varepsilon_{2}^{A}\varepsilon_{1}^{j}%
-C_{Aj}^{i}\varepsilon_{1}^{A}\varepsilon_{2}^{j}.
\end{gather}
Let us now introduce a new set of structure constants in terms of the
structure constants of the FDA:%
\begin{equation}
\tilde{C}_{IJ}^{K}=\left(
\begin{array}
[c]{cc}%
\tilde{C}_{BC}^{A} & \tilde{C}_{Aj}^{i}\\
\tilde{C}_{jA}^{i} & \tilde{C}_{jk}^{i}%
\end{array}
\right)  =\left(
\begin{array}
[c]{cc}%
C_{BC}^{A} & C_{Aj}^{i}\\
\left(  -1\right)  ^{p+1}C_{jA}^{i} & 0
\end{array}
\right)  . \label{Ctilde}%
\end{equation}
The new constants $\tilde{C}_{IJ}^{K}$ are written in terms of the generalized
index and it is straightforward to see that the products of the corresponding
$L_{\infty}$ algebra can be written in terms of them. Furthermore, by directly
calculating the Jacobiator, we find that the new structure constants satisfy
the standard Jacobi identity as a consequence of the standard and extended
Jacobi identities for $C_{BC}^{A}$ and $C_{Aj}^{i}$%
\begin{equation}
\tilde{C}_{I[J}^{K}\tilde{C}_{KL]}^{I}=0.
\end{equation}
This could be anticipated by the fact that the Antoniadis--Savvidy formulation
of Chern--Simons theories with $p$-forms from refs.
\cite{Antoniadis:2012ep,Antoniadis:2013jja}, consisting of the gauging of
high-degree differential forms in a Lie algebra, can be interpreted as the
gauging of a FDA1 whose $p$-form is introduced in the adjoint representation
of the Lie subalgebra, and that carries no cocycle representative of a
cohomology class. Thus, in this case, the FDA structure constants can be
codified into the structure constants of a Lie algebra by means of eq.
(\ref{Ctilde}). The definition of large gauge transformations immediately
allows us to extend the SW-GN formalism to include higher-degree differential
forms: given a stability subalgebra $\mathfrak{h}\subset\mathrm{FDA1}$, it is
possible to introduce the group element $z=e^{\xi}$ where $\xi=\xi^{I}t_{I}$
is a composite parameter evaluated in the coset space resulting from FDA1 and
the stability subalgebra. Then, we define the non-linear gauge field as the
following one-form%
\begin{equation}
\mu^{\xi}=\left(  \mu^{\xi}\right)  ^{A}t_{A}+\left(  \mu^{\xi}\right)
^{i}t_{i}=z^{-1}\mathrm{d}z+z^{-1}\mu z, \label{mutilde}%
\end{equation}
and, as it happens in the standard case, any functional of $\mu$ that is
invariant under the transformations of the stability subalgebra, will become
invariant under the transformations of the entire FDA1. In the case of
infinitesimal gauge parameters, eq. (\ref{mutilde}) reproduces eq.
(\ref{deltamufull}), and consequently, also eqs. (\ref{delta1}) and
(\ref{delta2}).

Let us consider the particular case of a FDA1 given by an extension of the
Poincar\'{e} algebra, where the representation index $i$ is chosen to be in
the adjoint representation and the differential degree of the higher-degree
forms is chosen as three. In this case, the composite gauge field is given by%
\begin{equation}
\mu=\mu^{A}t_{A}+\mu^{i}t_{i}.
\end{equation}
As we deal with a FDA1 extension of the Poincar\'{e} Lie algebra, the
algebraic index $A$ is decomposed as $A=\left(  a,ab\right)  $. Thus, the
component of the one-form are denoted by $\mu^{A}=\left(  e^{a},\omega
^{ab}\right)  $. In the same way, the vectors $t_{A}$ generating the
Poincar\'{e} Lie algebra are decomposed as $t_{A}=\left(  P_{a},J_{ab}\right)
$. The full $L_{\infty}$ algebra dual to the resulting FDA is composed by the
Poincar\'{e} subalgebra spanned by $\left\{  t_{A}\right\}  $, in addition to
a extended subspace spanned by the mappings $\left\{  t_{i}\right\}  $. Since
we have chosen the index $i$ to be the adjoint representation, it takes values
in the same domain as $A$. To avoid confussion with the vectors $t_{A}$ and
one-forms $\mu^{A}$ associated to the Lie subalgebra (or standard sector), we
denote the mathematical objects associated to the extended sector as
$t_{i}\equiv$ $\bar{t}_{A}=\left(  \bar{P}_{a},\bar{J}_{ab}\right)  $ and
$\mu^{i}\equiv$ $\bar{\mu}^{A}=\left(  b^{a},b^{ab}\right)  $\footnote{In the
following, we denote with a bar to the vectors lying in extended algebraic
sectors.}. Thus, the gauge fields corresponding to the Lie subalgebra and the
extended algebraic sector can be written as linear combinations of the vectors
spanning the $L_{\infty}$ algebra, as follows:
\begin{align}
\mu^{A}t_{A}  &  =e^{e}P_{a}+\frac{1}{2}\omega^{ab}J_{ab},\\
\mu^{i}t_{i}  &  \equiv\bar{\mu}^{A}\bar{t}_{A}=b^{a}\bar{P}_{a}+\frac{1}%
{2}b^{ab}\bar{J}_{ab}.
\end{align}
Here, $b^{a}$ and $b^{ab}$ are three-forms. According to eqs. (\ref{fda1A}%
)-(\ref{fda1C}), in this case the dual $L_{\infty}$ algebra spanned by the
complete set of vectors $\left\{  P_{a},J_{ab},\bar{P}_{a},\bar{J}%
_{ab}\right\}  $ take the following form%
\begin{align}
\left[  J_{ab},J_{cd}\right]   &  =\eta_{bc}J_{ad}+\eta_{ad}J_{bc}-\eta
_{bd}J_{ac}-\eta_{ac}J_{bd},\label{P1}\\
\left[  J_{ab},P_{c}\right]   &  =\eta_{bc}P_{a}-\eta_{ac}P_{b},\label{P2}\\
\left[  J_{ab},\bar{J}_{cd}\right]   &  =\eta_{bc}\bar{J}_{ad}+\eta_{ad}%
\bar{J}_{bc}-\eta_{bd}\bar{J}_{ac}-\eta_{ac}\bar{J}_{bd},\label{P3}\\
\left[  J_{ab},\bar{P}_{c}\right]   &  =\eta_{bc}\bar{P}_{a}-\eta_{ac}\bar
{P}_{b},\label{P4}\\
\left[  \bar{J}_{ab},P_{c}\right]   &  =\eta_{bc}\bar{P}_{a}-\eta_{ac}\bar
{P}_{b},\label{P5}\\
\text{others}  &  =0. \label{P5b}%
\end{align}
As it was shown in the example of section 2.1, we consider that the stability
subalgebra is the Lorentz algebra and introduce the following multiplet
evaluated in the remaining subspace%
\begin{equation}
\xi=\xi^{a}P_{a}+\zeta^{a}\bar{P}_{a}+\frac{1}{2}\zeta^{ab}\bar{J}_{ab},
\label{mult}%
\end{equation}
where $\xi^{a}$ is a zero-form, $\zeta^{a}$ and $\zeta^{ab}$ are two-forms,
and $\bar{P}_{a}$ and $\bar{J}_{ab}$ are operators defined according eq.
(\ref{identify}) carrying differential degree \thinspace$-2$. By directly
plugging in the exponential $z=e^{\xi}$ into eq. (\ref{mutilde}), we obtain
the following non-linear gauge field%
\begin{equation}
\mu^{\xi}=V^{a}P_{a}+\frac{1}{2}W^{ab}J_{ab}+B^{a}\bar{P}_{a}+\frac{1}%
{2}B^{ab}\bar{J}_{ab}, \label{tgf}%
\end{equation}
where each component is explicitly given by%
\begin{align}
V^{a}  &  =e^{a}+\mathrm{D}_{\omega}\xi^{a},\\
W^{ab}  &  =\omega^{ab},\\
B^{a}  &  =b^{a}+\mathrm{D}_{\omega}\zeta^{a}+\frac{1}{2}\mathrm{D}_{\omega
}\zeta^{ab}\xi_{b}-\frac{1}{2}\zeta_{\text{ \ }c}^{a}\mathrm{D}_{\omega}%
\xi^{c},\\
B^{ab}  &  =b^{ab}+\mathrm{D}_{\omega}\zeta^{ab}.
\end{align}
Although we have obtained the explicit form of the non-linear fields, it is
still necessary to specify the transformation law of the multiplet $\xi$. By
considering an infinitesimal gauge transformation with a parameter evaluated
in the Lorentz algebra%
\begin{equation}
\varepsilon=\frac{1}{2}\varepsilon^{ab}J_{ab}.
\end{equation}
The multiplet transforms linearly, according eq. (\ref{5Binf}), as follows%
\begin{align}
\delta\xi^{a}  &  =-\varepsilon_{\text{ \ }c}^{a}\xi^{c},\\
\delta\zeta^{a}  &  =-\varepsilon_{\text{ \ }c}^{a}\zeta^{c},\\
\delta\zeta^{ab}  &  =-\varepsilon_{\text{ \ }c}^{a}\zeta^{cb}+\varepsilon
_{\text{ \ }c}^{b}\zeta^{ca}.
\end{align}
On the other hand, if we consider a gauge transformation along the coset space
generators, with a gauge parameter denoted by%
\begin{equation}
\varepsilon=\varepsilon^{a}P_{a}+\bar{\varepsilon}^{a}\bar{P}_{a}+\frac{1}%
{2}\bar{\varepsilon}^{ab}\bar{J}_{ab}, \label{eps}%
\end{equation}
then, the multiplet transforms according eq. (\ref{5Cinf}). Thus, by plugging
in eqs. (\ref{mult}) and (\ref{eps}) and applying the algebra relations, one
finds that the components of $\xi$ transform as%
\begin{align}
\delta\xi^{a}  &  =-\varepsilon^{a},\\
\delta\zeta^{a}  &  =-\bar{\varepsilon}^{a}+\frac{1}{2}\zeta_{\text{ \ }c}%
^{a}\varepsilon^{c}-\frac{1}{2}\bar{\varepsilon}_{\text{ \ }c}^{a}\xi^{c},\\
\delta\zeta^{ab}  &  =-\bar{\varepsilon}^{ab}.
\end{align}
Thus, we have introduced a non-linear realization for the FDA. Any Lorentz
invariant Lagrangian constructed with $\mu^{\xi}$ instead of $\mu$ will
immediately become fully gauge invariant under the transformations of the
FDA1. The original non-invariance under the transformations of the coset space
is compensated by the transformation law of $\xi$. This completes the gauging
of the FDA1 with trivial cohomology in the SW-GN formalism and allows to
formulate Poincar\'{e} gauge invariant theories in arbitrary dimensions while
coupling a three-form in the adjoint representation of the gauge group.

\subsection{Non-trivial cohomology}

As in the previous case, in order to introduce a non-linear gauge field it is
necessary to define large gauge transformations. However, when the extended
Maurer--Cartan equations carry non-trivial cocycles, the exponentiation of the
gauge parameters cannot be performed as well. This is due to the presence of a
multilinear product in the transformation law of the extended field. It is
therefore necessary a definition of large gauge transformation that includes
the FDA1 cocycle. We overcome this obstacle by using a definition of large
gauge transformation that makes no use of an exponential. For this purpose, we
first study the composition law of infinitesimal gauge transformations in the
context of FDA1 gauge theory. Let us consider two independent gauge
transformations along FDA1 gauge parameters $\varepsilon_{1}$ and
$\varepsilon_{2}$. A direct calculation shows that their commutator is given
by a third gauge transformation:%
\begin{equation}
\left[  \delta_{\varepsilon_{1}},\delta_{\varepsilon_{2}}\right]  \mu
=\delta_{\varepsilon_{3}}\mu,
\end{equation}
where the composite parameter $\varepsilon_{3}=\varepsilon_{3}^{A}%
t_{A}+\varepsilon_{3}^{i}t_{i}$ is given in terms of the $L_{\infty}$ products
by%
\begin{equation}
\varepsilon_{3}=\left[  \varepsilon_{1},\varepsilon_{2}\right]  _{2}+\frac
{1}{p!\left(  p-1\right)  !}\left[  \varepsilon_{1},\varepsilon_{2},\mu
^{p-1}\right]  _{p+1},
\end{equation}
or, equivalently, in terms of the FDA1 structure constants by%
\begin{align}
\varepsilon_{3}^{A} &  =C_{BC}^{A}\varepsilon_{1}^{B}\varepsilon_{2}^{C},\\
\varepsilon_{3}^{i} &  =C_{Aj}^{i}\left(  \varepsilon_{1}^{A}\varepsilon
_{2}^{j}-\varepsilon_{2}^{A}\varepsilon_{1}^{j}\right)  +\frac{1}{\left(
p-1\right)  !}C_{A_{1}\cdots A_{p+1}}^{i}\varepsilon_{1}^{A_{1}}%
\varepsilon_{2}^{A_{2}}\mu^{A_{3}}\cdots\mu^{A_{p+1}}.
\end{align}
The composition law of the gauge parameters induces a definition of
generalized bilinear product for FDA1 as a shifting of the bilinear product
$\left[  ,\right]  _{2}$ of the Lie subalgebra%
\begin{equation}
\left[  \varepsilon_{1},\varepsilon_{2}\right]  \equiv\left[  \varepsilon
_{1},\varepsilon_{2}\right]  _{2}+\frac{1}{p!\left(  p-1\right)  !}\left[
\varepsilon_{1},\varepsilon_{2},\mu^{p-1}\right]  _{p+1}.\label{def2}%
\end{equation}
This definition allows us to write both the parameter composition law and the
transformation law of the FDA1 curvature in a more convenient way:%
\begin{equation}
\varepsilon_{3}=\left[  \varepsilon_{1},\varepsilon_{2}\right]  ,\text{
\ \ \ \ }\delta R=\left[  R,\varepsilon\right]  .
\end{equation}
The notion of a bilinear product carrying the information of the FDA1 products
allows us to generalize the definition of large gauge transformations of Lie
algebras given in eq. (\ref{large1}). We define the large FDA1 gauge
transformations as%
\begin{align}
\delta_{\text{large}}\mu &  =\nabla\varepsilon-\frac{1}{2}\left[
\varepsilon,\nabla\varepsilon\right]  +\frac{1}{3!}\left[  \varepsilon,\left[
\varepsilon,\nabla\varepsilon\right]  \right]  +\cdots\nonumber\\
&  =\sum_{n=1}^{\infty}\frac{\left(  -1\right)  ^{n+1}}{n!}\Delta^{n}%
\mu,\label{fdalarge}%
\end{align}
where, for convenience, we denote%
\begin{equation}
\Delta^{n}\mu=\left[  \varepsilon,\overset{n-1}{\ldots},\left[  \varepsilon
,\nabla\varepsilon\right]  \right]  ,
\end{equation}
with the convention $\Delta^{1}\mu=\nabla\varepsilon$ and $\Delta
\varepsilon=0$. Eq. (\ref{fdalarge}) makes use of the generalized bilinear
bracket of FDA1 defined in eq. (\ref{def2}). Thus, $\Delta^{n}\mu$ represents
the contribution of degree $n$ in $\varepsilon$ to the large gauge variation
of $\mu$. As will be shown further on, it is possible to find explicit
expressions of large gauge transformations in the context of particular cases
of FDA1 by directly calculating each term in the sum of eq. (\ref{large1}).
The definition in eq. (\ref{fdalarge}) is equivalent to eq. (\ref{large1}) in
the case of the Lie algebra. However, it does not make use of a group element.
Thus, eq. (\ref{fdalarge}) allows to analytically extend the large gauge
transformations of a Lie algebra to the case of a general FDA1, in a way that
include the corrections coming from the FDA1 cocycle. A transformed gauge
field can be then explicitly calculated as long as the sum in the r.h.s. of
eq. (\ref{fdalarge}) is truncated at some degree. Furthermore, this definition
allows to introduce a non-linear gauge field $\mu^{\xi}$ with $\xi$ a FDA1
multiplet evaluated in the coset space resulting from the FDA1 and a chosen
stability subalgebra $\mathfrak{h}$%
\begin{equation}
\xi\in\frac{\text{\textrm{FDA1}}}{\mathfrak{h}}.
\end{equation}
As it happens in the SW-GN formalism for Lie algebras, the multiplet $\xi$ is
considered as a fundamental field of the gauge theory and not as the parameter
of a symmetry transformation. It is therefore necessary to define a
transformation law for $\xi$ that is consistent with the definition of the
non-linear gauge field. If the gauge algebra is a Lie algebra, the
infinitesimal variations of the zero-form under the stability subalgebra and
the coset space are given by eqs. (\ref{5Binf}) and (\ref{5Cinf})
respectively. As it happens with the gauge field, it is possible to find
analytic extensions of such transformation laws in a way that includes the
cocycle of the FDA1 and that is consistent with the large gauge transformation
introduced in eq. (\ref{fdalarge}). Let us now consider an infinitesimal gauge
parameter $\varepsilon$ evaluated in the stability subalgebra of FDA1. We
extend the transformation law from eq. (\ref{5Binf}), by plugging in the
shifting of eq. (\ref{def2}). Thus, the infinitesimal gauge variation of the
multiplet becomes%
\begin{equation}
\delta\xi=\left[  \xi,\varepsilon\right]  \equiv\left[  \xi,\varepsilon
\right]  _{2}+\frac{1}{p!\left(  p-1\right)  !}\left[  \xi,\varepsilon
,\mu^{p-1}\right]  _{p+1}.\label{fdalinear}%
\end{equation}
Similarly, for an infinitesimal gauge parameter lying in the coset space, we
consider the analytic extension of eq. (\ref{5Cinf}) by plugging in the
shifting of eq. (\ref{def2}). In this case, eq. (\ref{5Cinf}) implies that the
gauge variation of the FDA1 multiplet $\delta\xi$ must verify the following
condition:%
\begin{equation}
\left.  \sum_{n=0}^{\infty}\left[  \frac{\left(  -1\right)  ^{n}}{n!}\left[
\xi,\overset{n}{\cdots},\left[  \xi,\varepsilon\right]  \cdots\right]
+\frac{\left(  -1\right)  ^{n}}{\left(  n+1\right)  !}\left[  \xi
,\overset{n}{\cdots},\left[  \xi,\delta\xi\right]  \cdots\right]  \right]
\right\vert _{\text{\textrm{FDA1/}}\mathfrak{h}}=0,\label{hh}%
\end{equation}
where the bracket $\left[  ,\right]  $ denotes the product defined in eq.
(\ref{def2}). This condition is equivalent to the statement that the sum in
the l.h.s of eq. (\ref{hh}) must be a vector of the stability algebra, while
its component along the coset subspace vanishes. The vanishing of the sum in
the coset space allows to explicitly obtain the transformation law of $\xi$
for particular cases just as it happens in the Lie algebra case. Thus, by
fixing the transformation law of $\xi$ according to (\ref{hh}), the non-linear
gauge field transforms as a connection of the stability subgroup, making
possible the formulation of gauge invariant theories for FDA1.

\section{Bosonic FDA gauge theories}

\subsection{Poincar\'{e}-FDA gauge theory}

Let us now study the particular case of a FDA1 constructed as a FDA1 extension
of the bosonic $D$-dimensional Poincar\'{e} algebra. The Maurer--Cartan
potentials are given by a one-form, decomposed as $\mu^{A}=\left(
e^{a},\omega^{ab}\right)  $ and a three-form. We choose the three-form to be
in the adjoint representation of the Lie algebra described by $\mu^{A}$.
Therefore, it is also decomposed as $\mu^{i}=\left(  b^{a},b^{ab}\right)  $.
To choose the adjoint representation means that the first set of generalized
structure constants become the structure constants of the Lie algebra, i.e.,
$C_{Aj}^{i}\rightarrow C_{AB}^{C}$. Similarly, the structure constants of the
cocycle become $C_{A_{1}\cdots A_{p+1}}^{i}\rightarrow C_{A_{1}\cdots A_{p+1}%
}^{A}$. As we did in section 3, we rename the dual vectors associated to the
extended directions as $t_{i}\rightarrow\bar{t}_{A}$. The resulting FDA is
given by eq. (\ref{mu}), in addition to eq. (\ref{Bv2}). We construct the
algebra by including the following cocycle, which is not covariatly exact, and
therefore representative of a particular Chevalley--Eilenberg cohomology
class:%
\begin{equation}
\Omega^{A}=\left(  e^{a}\omega^{bc}\omega_{bd}\omega_{\text{ \ }c}%
^{d},0\right)  .
\end{equation}
It is easy to verify the closure of $\Omega^{A}$ under the covariant
derivative of the Poincar\'{e} algebra. Thus, the Poincar\'{e}-FDA is
described by the following Maurer--Cartan equations%
\begin{align}
T^{a}  &  =\mathrm{d}e^{a}+\omega_{\text{ \ }b}^{a}e^{b}=0,\label{mcp1}\\
R^{ab}  &  =\mathrm{d}\omega^{ab}+\omega_{\text{ \ }c}^{a}\omega
^{cb}=0,\label{mcp2}\\
H^{a}  &  =\mathrm{d}b^{a}+\omega_{\text{ \ }c}^{a}b^{c}+b_{\text{ \ }c}%
^{a}e^{c}+e^{a}\omega^{bc}\omega_{bd}\omega_{\text{ \ }c}^{d}=0,\label{mcp3}\\
H^{ab}  &  =\mathrm{d}b^{ab}+\omega_{\text{ \ }c}^{a}b^{cb}-\omega_{\text{
\ }c}^{b}b^{ca}=0. \label{mcp4}%
\end{align}
We denote the dual vectors as $t_{A}=\left(  P_{a},J_{ab}\right)  $ for the
standard sector and and $\bar{t}_{A}=\left(  \bar{P}_{a},\bar{J}_{ab}\right)
$ for the extended one. By extracting the information about the cocycle
structure constants, we can write down products that define the dual algebra,
that we denote by $L_{\infty}^{\text{\textrm{Poincar\'{e}-FDA}}}$, which are
given by eqs. (\ref{P1})-(\ref{P5b}) in addition to the following multilinear
product:%
\begin{align}
\left[  P_{a},J_{bc},J_{de},J_{fg}\right]   &  =\left(  3!\right)  ^{2}\left(
\eta_{bd}\eta_{cg}\eta_{ef}-\eta_{cd}\eta_{bg}\eta_{ef}-\eta_{be}\eta_{cg}%
\eta_{df}+\eta_{ce}\eta_{bg}\eta_{df}\right. \nonumber\\
&  \left.  -\eta_{bd}\eta_{cf}\eta_{eg}+\eta_{cd}\eta_{bf}\eta_{eg}+\eta
_{be}\eta_{cf}\eta_{dg}-\eta_{ce}\eta_{bf}\eta_{dg}\right)  \bar{P}_{a},
\label{P6}%
\end{align}

With the purpose of writing a non-linear realization of Poincar\'{e}-FDA, we
gauge the algebra by introducing a composite gauge field carrying components
in both algebraic sectors. According to (\ref{one}), we can use the dual basis
of vectors and write the gauge fields as follows%
\begin{equation}
\mu=e^{a}P_{a}+\frac{1}{2}\omega^{ab}J_{ab}+b^{a}\bar{P}_{a}+\frac{1}{2}%
b^{ab}\bar{J}_{ab},
\end{equation}
where the components along $\left\{  P_{a},J_{ab}\right\}  $ and $\left\{
\bar{P}_{a},\bar{J}_{ab}\right\}  $ are evaluated in the standard and extended
sectors of the Poincar\'{e}-FDA respectively. Moreover, we introduce a
composite gauge parameter $\varepsilon$ carrying the following standard and
extended components%
\begin{equation}
\varepsilon=\varepsilon^{a}P_{a}+\frac{1}{2}\varepsilon^{ab}J_{ab}%
+\bar{\varepsilon}^{a}\bar{P}_{a}+\frac{1}{2}\bar{\varepsilon}^{ab}\bar
{J}_{ab},
\end{equation}
We now plug in eqs. (\ref{P1})-(\ref{P5b}) and (\ref{P6}) into to eq.
(\ref{deltamualone}), and set $p=3$, to calculate the gauge variation of the
gauge field. By making use of the dual basis $\left\{  t_{A},\bar{t}%
_{A}\right\}  $ and the relations of $L_{\infty}%
^{\text{\textrm{Poincar\'{e}-FDA}}}$, we express the infinitesimal gauge
variation in terms of a single mathematical object without separating in
standard and extended sectors. Thus, the gauge variation can be expressed as%
\begin{align}
\delta\mu &  =\left(  \mathrm{d}\varepsilon^{a}+\omega_{\text{ \ }c}%
^{a}\varepsilon^{c}-\varepsilon_{\text{ \ }c}^{a}e^{c}\right)  P_{a}+\frac
{1}{2}\left(  \mathrm{d}\varepsilon^{ab}+2\omega_{\text{ \ }c}^{a}%
\varepsilon^{cb}\right)  J_{ab}\nonumber\\
&  +\left(  \mathrm{d}\bar{\varepsilon}^{a}+\omega_{\text{ \ }c}^{a}%
\bar{\varepsilon}^{c}+b_{\text{ \ }c}^{a}\varepsilon^{c}-\varepsilon_{\text{
\ }c}^{a}b^{c}-\bar{\varepsilon}_{\text{ \ }c}^{a}e^{c}+\varepsilon^{a}%
\omega^{bc}\omega_{c}^{\text{ \ }f}\omega_{fb}+3e^{a}\varepsilon_{bc}%
\omega_{\text{ \ }d}^{b}\omega^{dc}\right)  \bar{P}_{a}\nonumber\\
&  +\frac{1}{2}\left(  \mathrm{d}\bar{\varepsilon}^{ab}+2\omega^{ac}%
\bar{\varepsilon}_{c}^{\text{ \ }b}-2\varepsilon_{\text{ \ }c}^{a}%
b^{cb}\right)  \bar{J}_{ab},
\end{align}
or, equivalently, in components as%
\begin{align}
\delta e^{a}  &  =\mathrm{D}_{\omega}\varepsilon^{a}-\varepsilon_{\text{ \ }%
c}^{a}e^{c},\label{dP1}\\
\delta\omega^{ab}  &  =\mathrm{D}_{\omega}\varepsilon^{ab},\label{dP2}\\
\delta b^{a}  &  =\mathrm{D}_{\omega}\bar{\varepsilon}^{a}+b_{\text{ \ }c}%
^{a}\varepsilon^{c}-\varepsilon_{\text{ \ }c}^{a}b^{c}-\bar{\varepsilon
}_{\text{ \ }c}^{a}e^{c}+\varepsilon^{a}\omega^{bc}\omega_{c}^{\text{ \ }%
f}\omega_{fb}+3e^{a}\varepsilon_{bc}\omega_{\text{ \ }d}^{b}\omega
^{dc},\label{dP3}\\
\delta b^{ab}  &  =\mathrm{D}_{\omega}\bar{\varepsilon}^{ab}-\varepsilon
_{\text{ \ }c}^{a}b^{cb}+\varepsilon_{\text{ \ }c}^{b}b^{ca}. \label{dP4}%
\end{align}
Notice that the structure constants of the cocycle are present in the last two
terms of the r.h.s. of eq. (\ref{dP3}).

Let us now introduce a large transformation performed with respect to a finite
parameter that lies in the coset space between the full Poincar\'{e}-FDA and a
stability subalgebra that we choose as the Lorentz algebra spanned by
$\left\{  J_{ab}\right\}  $%
\begin{equation}
\xi=\xi^{a}P_{a}+\zeta^{a}\bar{P}_{a}+\frac{1}{2}\zeta^{ab}\bar{J}_{ab}%
\in\frac{\text{\textrm{Poincar\'{e}-FDA}}}{\text{\textrm{SO}}\left(
d-1,1\right)  }.
\end{equation}
Since we do not consider a parametrization for the Lorentz subalgebra, the
linear contributions to the large gauge transformations, denoted by $\Delta
\mu$, have the same functional form that the ones in eqs. (\ref{dP1}%
)-(\ref{dP4}) with $\rho^{ab}=0$. Thus, $\Delta\mu$ is given in components by%
\begin{align}
\Delta\mu &  =\left(  \mathrm{d}\xi^{a}+\omega_{\text{ \ }c}^{a}\xi
^{c}\right)  P_{a}+\left(  \mathrm{d}\zeta^{a}+\omega_{\text{ \ }c}^{a}%
\zeta^{c}+b_{\text{ \ }c}^{a}\xi^{c}-\zeta_{\text{ \ }c}^{a}e^{c}+\xi
^{a}\omega^{bc}\omega_{c}^{\text{ \ }f}\omega_{fb}\right)  \bar{P}%
_{a}\nonumber\\
&  +\frac{1}{2}\left(  \mathrm{d}\zeta^{ab}+2\omega_{\text{ \ }c}^{a}%
\zeta^{cb}\right)  \bar{J}_{ab}.
\end{align}
By directly calculating the higher-degree variations of the gauge field, we
find that the sum of eq. (\ref{fdalarge}) is truncated, which allows us to
write down the large gauge transformation. We denote its components as
\begin{equation}
\mu^{\xi}=V^{a}P_{a}+\frac{1}{2}W^{ab}J_{ab}+B^{a}\bar{P}_{a}+\frac{1}%
{2}B^{ab}\bar{J}_{ab},
\end{equation}
where each one is explicitly given by%
\begin{align}
V^{a}  &  =e^{a}+\mathrm{D}_{\omega}\xi^{a},\label{V}\\
W^{ab}  &  =\omega^{ab},\label{W}\\
B^{a}  &  =b^{a}+\mathrm{D}_{\omega}\zeta^{a}+\frac{1}{2}\mathrm{D}_{\omega
}\zeta^{ab}\rho_{b}-\frac{1}{2}\zeta_{\text{ \ }c}^{a}\mathrm{D}_{\omega}%
\xi^{c}+b_{\text{ \ }c}^{a}\xi^{c}-\zeta_{\text{ \ }c}^{a}e^{c}+\xi^{a}%
\omega^{bc}\omega_{c}^{\text{ \ }f}\omega_{fb},\label{B1}\\
B^{ab}  &  =b^{ab}+\mathrm{D}_{\omega}\zeta^{ab}. \label{B2}%
\end{align}
Notice that, as it happens with the infinitesimal transformation, the
information about the cocycle is only present in the definition of the
three-form $B^{a}$, which is also the only one that contains non-linear terms
in the parameters of the transformation. Similarly, the cocycle structure
constants are also present in the transformation law of the two-form multiplet
$\zeta^{a}$ associated to the same dual vector. The transformation law of the
Poincar\'{e}-FDA multiplet $\xi$ can be found in appendix \ref{app1}. Eqs.
(\ref{V}) and (\ref{W}) reproduce the non-linear gauge field of the
Poincar\'{e} Lie algebra. This is a natural consequence of the presence of the
Poincar\'{e} algebra as subalgebra of Poincar\'{e}-FDA. Thus, an action
principle constructed as a functional of $\mu^{\xi}$ must not only preserve
the invariance under the Poincar\'{e} transformations but also extend it along
the extended directions described by higher-degree parameters.

In order to construct an action principle let us set $D=5$ and consider
non-vanishing curvature in eqs. (\ref{mcp1})-(\ref{mcp4}). This implies that
the curvature associated to the non-linear gauge field is also non-vanishing.
We denote the two-forms and four-forms associated to $\mu^{\xi}$
as\footnote{We do not include the Lorentz curvature since it is the same that
appears in eq. (\ref{mcp2}). It remains invariant under the gauge
transformation as well as $\omega^{ab}$ does.}:%
\begin{align}
\mathcal{T}^{a}  &  =\mathrm{D}_{\omega}V^{a},\\
\mathcal{H}^{a}  &  =\mathrm{D}_{\omega}B^{a}+B_{\text{ \ }c}^{a}V^{c}%
+V^{a}\omega^{bc}\omega_{bd}\omega_{\text{ \ }c}^{d},\\
\mathcal{H}^{ab}  &  =\mathrm{D}_{\omega}B^{ab}.
\end{align}
A five-dimensional Lorentz-invariant modification to the Einstein--Hilbert
Lagrangian that couples the three-form is given by the following action
principle%
\begin{equation}
I=\kappa\int\left(  \epsilon_{abcde}R^{ab}e^{c}e^{d}e^{e}+H^{a}e_{a}\right)  ,
\end{equation}
with $\kappa$ a dimensional constant. This action principle becomes gauge
invariant under Poincar\'{e}-FDA by considering the non-linear gauge fields,
i.e.,%
\begin{equation}
I^{\mathrm{Poincar\acute{e}}\text{\textrm{-}}\mathrm{FDA}}=\kappa\int\left(
\epsilon_{abcde}R^{ab}V^{c}V^{d}V^{e}+\mathcal{H}^{a}V_{a}\right)  .
\label{action1}%
\end{equation}
In general, any Lorentz invariant Lagrangian depending on the gauge fields and
curvatures of Poincar\'{e}-FDA becomes fully gauge invariant by considering
the transformed gauge fields as building blocks. Notice that, although the
action principle extends five-dimensional general relativity by coupling
three-form, it does not involve the cocycle in the Lagrangian density, since
its contribution to $\mathcal{H}^{a}$ vanishes when contracting with $V^{a}$
in the second term inside the integral of the r.h.s. of eq. (\ref{action1}).
This feature is specific to the five-dimensional case. As it will be shown in
the following section, in higher dimensions, action principles constructed by
this process in general do couple the cocycle in the Lagrangian density and
the corresponding equations of motion. However, for the purposes of this
example we will only consider the equations of motion emerging form eq.
(\ref{action1}). We interpret the terms proportional to the Einstein--Hilbert
Lagrangian and the one depending on the three-forms as pure gravity and matter
Lagrangians respectively. Thus, we can write the action principle as%
\begin{equation}
I^{\mathrm{Poincar\acute{e}}\text{\textrm{-}}\mathrm{FDA}}=\kappa\int\left(
L_{\text{\textrm{G}}}+L_{\text{\textrm{M}}}\right)  . \label{action1b}%
\end{equation}
We now introduce a spin and energy-momentum forms $\sigma_{ab}$ and $\tau_{a}$
in terms of the functional variations of the matter Lagrangian with respect to
the geometric fields of the theory%
\begin{align}
\delta_{V}L_{\text{\textrm{M}}}  &  =-\delta V^{a}\ast\tau_{a}%
,\label{emtensor}\\
\delta_{\omega}L_{\text{\textrm{M}}}  &  =-\delta\omega^{ab}\ast\sigma_{ab},
\label{spin}%
\end{align}
where $\ast$ denotes the Hodge dual operator mapping $q$-forms into $\left(
5-q\right)  $-forms. Thus, the gravity equations of motion are given by%
\begin{align}
\varepsilon_{a}  &  \equiv\frac{\delta L_{\text{\textrm{G}}}}{\delta V^{a}%
}=3\epsilon_{abcde}R^{bc}V^{d}V^{e}=-\ast\tau_{a},\label{e1}\\
\varepsilon_{ab}  &  \equiv\frac{\delta L_{\text{\textrm{G}}}}{\delta
\omega^{ab}}=3\epsilon_{abcde}\mathcal{T}^{c}V^{d}V^{e}=-\ast\sigma_{ab}.
\label{e2}%
\end{align}
with
\begin{align}
\ast\tau_{a}  &  =-\mathrm{D}_{\omega}B_{a}-2B_{ab}V^{b},\\
\ast\sigma_{ab}  &  =\frac{1}{2}\left(  B_{a}V_{b}-B_{b}V_{a}\right)  .
\end{align}
Notice that, in absence of three-forms, the equations of motion become
$\varepsilon_{a}=0$ and $\varepsilon_{ab}=0$, thus reproducing the standard
general relativity dynamics in the Cartan formalism and presenting on-shell
vanishing torsion as a consequence of eq. (\ref{e2}). In contrast, the
presence of three-forms generate non-vanishing torsion and a modified dynamics
for both the vielbein and spin connection.

\subsection{Maxwell-FDA gauge theory}

Let us now consider an extension of the bosonic Maxwell algebra. The
left-invariant one-forms are, in the chosen basis, split as $\mu^{A}=\left(
e^{a},\omega^{ab},k^{ab}\right)  $. The Maxwell Lie algebra is then defined by
the following Maurer--Cartan equations
\cite{Bacry:1968zf,Bacry:1970ye,Schrader:1972zd}
\begin{align}
T^{a}  &  =\mathrm{d}e^{a}+\omega_{\text{ \ }c}^{a}e^{c}=0,\label{mfda1a}\\
R^{ab}  &  =\mathrm{d}\omega^{ab}+\omega_{\text{ \ }c}^{a}\omega
^{cb}=0,\label{mfda1b}\\
F^{ab}  &  =\mathrm{d}k^{ab}+\omega_{\text{ \ }c}^{a}k^{cb}-\omega_{\text{
\ }c}^{b}k^{ca}+\frac{1}{l^{2}}e^{a}e^{b}=0, \label{mfda1c}%
\end{align}
i.e., by the zero-cuvature statement $R^{A}=\left(  T^{a},R^{ab}%
,F^{ab}\right)  =\left(  0,0,0\right)  $. Similarly to the previous case, we
introduce a three-form in the adjoint representation of the Lie algebra, to
whose components we denote by $\left(  b^{a},b^{ab},\beta^{ab}\right)  $, and
extend the Lie algebra to a FDA1 through the inclusion of a non-trivial
four-cocycle representative of a Chevalley--Eilenberg cohomology class in the
adjoint representation of the Maxwell algebra. That cocycle, denoted by
$\Omega^{A}$, is explicitly given by \cite{Salgado:2021eli}%
\begin{align}
\Omega^{A}  &  =\left(  ~0,~0,~\Omega^{ab}\right)  ,\\
\Omega^{ab}  &  =~k_{\text{ \ }c}^{a}k_{\text{ \ }d}^{c}e^{d}e^{b}-k_{\text{
\ }c}^{b}k_{\text{ \ }d}^{c}e^{d}e^{a}-2k^{ab}k_{cd}e^{c}e^{d}~.
\end{align}
Notice that the only non-vanishing component is present in the third algebraic
sector which, in the case of the one-forms, corresponds to the field $k^{ab}$.
We define a new set of Maurer--Cartan equations for the three-forms, and
introduce the cocycle according to eq. (\ref{Bv2}):
\begin{align}
H^{a}  &  =\mathrm{d}b^{a}+\omega_{\text{ \ }c}^{a}b^{c}+b_{\text{ \ }c}%
^{a}e^{c}=0,\label{mfda1d}\\
H^{ab}  &  =\mathrm{d}b^{ab}+\omega_{\text{ \ }c}^{a}b^{bc}-\omega_{\text{
\ }c}^{b}b^{ba}=0,\label{mfda1e}\\
G^{ab}  &  =\mathrm{d}\beta^{ab}+\omega_{\text{ \ }c}^{a}\beta^{cb}%
-\omega_{\text{ \ }c}^{b}\beta^{ca}+b_{\text{ \ }c}^{a}k^{cb}-b_{\text{ \ }%
c}^{b}k^{ca}+\frac{1}{l^{2}}\left(  e^{a}b^{b}-e^{b}b^{a}\right)  +\Omega
^{ab}=0. \label{mfda1f}%
\end{align}
Thus, eqs. (\ref{mfda1a})-(\ref{mfda1c}) in addition to eqs. (\ref{mfda1d}%
)-(\ref{mfda1f}) define a particular FDA1 to which we refer as Maxwell-FDA. As
it happens in the section 4.1, it is possible to extract the information of
the structure constants from eqs. (\ref{mfda1a})-(\ref{mfda1f}) and write down
the corresponding dual algebra in terms of multilinear products, which we
denote as $L_{\infty}^{\text{\textrm{Maxwell-FDA}}}$:

Bilinear products%
\begin{align}
\left[  J_{ab},J_{cd}\right]   &  =\eta_{bc}J_{ad}+\eta_{ad}J_{bc}-\eta
_{bd}J_{ac}-\eta_{ac}J_{bd},\\
\left[  J_{ab},P_{c}\right]   &  =\eta_{bc}P_{a}-\eta_{ac}P_{b},\\
\left[  P_{a},P_{b}\right]   &  =Z_{ab},\\
\left[  J_{ab},Z_{cd}\right]   &  =\eta_{bc}Z_{ad}+\eta_{ad}Z_{bc}-\eta
_{bd}Z_{ac}-\eta_{ac}Z_{bd},\\
\left[  J_{ab},\bar{J}_{cd}\right]   &  =\eta_{bc}\bar{J}_{ad}+\eta_{ad}%
\bar{J}_{bc}-\eta_{bd}\bar{J}_{ac}-\eta_{ac}\bar{J}_{bd},\\
\left[  J_{ab},\bar{P}_{c}\right]   &  =\eta_{bc}\bar{P}_{a}-\eta_{ac}\bar
{P}_{b},\\
\left[  J_{ab},\bar{Z}_{cd}\right]   &  =\eta_{bc}\bar{Z}_{ad}+\eta_{ad}%
\bar{Z}_{bc}-\eta_{bd}\bar{Z}_{ac}-\eta_{ac}\bar{Z}_{bd},\\
\left[  \bar{J}_{ab},P_{c}\right]   &  =\eta_{bc}\bar{P}_{a}-\eta_{ac}\bar
{P}_{b},\\
\left[  P_{a},\bar{P}_{b}\right]   &  =\bar{Z}_{ab},\\
\left[  \bar{J}_{ab},Z_{cd}\right]   &  =\eta_{bc}\bar{Z}_{ad}+\eta_{ad}%
\bar{Z}_{bc}-\eta_{bd}\bar{Z}_{ac}-\eta_{ac}\bar{Z}_{bd}.
\end{align}

Multilinear product%
\begin{align}
\left[  Z_{cd},Z_{ef},P_{g},P_{h}\right]  _{4}  &  =6\left(  \delta_{c}%
^{a}\delta_{h}^{b}\eta_{de}\eta_{fg}-4\delta_{c}^{a}\delta_{d}^{b}\eta
_{eg}\eta_{fh}-\delta_{d}^{a}\delta_{h}^{b}\eta_{ce}\eta_{fg}\right.
\nonumber\\
&  -\delta_{c}^{a}\delta_{h}^{b}\eta_{df}\eta_{eg}+4\delta_{c}^{a}\delta
_{d}^{b}\eta_{fg}\eta_{eh}+\delta_{d}^{a}\delta_{h}^{b}\eta_{cf}\eta
_{eg}\nonumber\\
&  -\delta_{e}^{a}\delta_{h}^{b}\eta_{fc}\eta_{dg}+4\delta_{e}^{a}\delta
_{f}^{b}\eta_{cg}\eta_{dh}+\delta_{f}^{a}\delta_{h}^{b}\eta_{ec}\eta
_{dg}\nonumber\\
&  +\delta_{e}^{a}\delta_{h}^{b}\eta_{fd}\eta_{cg}-4\delta_{e}^{a}\delta
_{f}^{b}\eta_{dg}\eta_{ch}-\delta_{f}^{a}\delta_{h}^{b}\eta_{ed}\eta
_{cg}\nonumber\\
&  -\delta_{c}^{a}\delta_{g}^{b}\eta_{de}\eta_{fh}+\delta_{d}^{a}\delta
_{g}^{b}\eta_{ce}\eta_{fh}+\delta_{c}^{a}\delta_{g}^{b}\eta_{df}\eta
_{eh}-\delta_{d}^{a}\delta_{g}^{b}\eta_{cf}\eta_{eh}\nonumber\\
&  \left.  +\delta_{e}^{a}\delta_{g}^{b}\eta_{fc}\eta_{dh}-\delta_{f}%
^{a}\delta_{g}^{b}\eta_{ec}\eta_{dh}-\delta_{e}^{a}\delta_{g}^{b}\eta_{fd}%
\eta_{ch}+\delta_{f}^{a}\delta_{g}^{b}\eta_{ed}\eta_{ch}\right)  \bar{Z}_{ab},
\end{align}%
\begin{equation}
\text{others}=0.
\end{equation}

We now gauge the algebra by introducing non-zero curvature. Moreover, we use
the dual basis to write down the gauge field $\mu$ and the gauge parameter
$\varepsilon$ as linear combinations of the vectors of the standard and
extended sectors:%
\begin{align}
\mu &  =e^{a}P_{a}+\frac{1}{2}\omega^{ab}J_{ab}+\frac{1}{2}k^{ab}Z_{ab}%
+b^{a}\bar{P}_{a}+\frac{1}{2}b^{ab}\bar{J}_{ab}+\frac{1}{2}\beta^{ab}\bar
{Z}_{ab},\\
\varepsilon &  =\varepsilon^{a}P_{a}+\frac{1}{2}\varepsilon^{ab}J_{ab}%
+\frac{1}{2}\epsilon^{ab}Z_{ab}+\bar{\varepsilon}^{a}\bar{P}_{a}+\frac{1}%
{2}\bar{\varepsilon}^{ab}\bar{J}_{ab}+\frac{1}{2}\bar{\epsilon}^{ab}\bar
{Z}_{ab}.
\end{align}
By plugging in the relations of $L_{\infty}^{\text{\textrm{Maxwell-FDA}}}$
into eq. (\ref{deltamualone}), we find the following infinitesimal gauge
variations for the components of the gauge field%
\begin{align}
\delta e^{a}  &  =\mathrm{D}_{\omega}\varepsilon^{a}-\varepsilon_{\text{ \ }%
c}^{a}e^{c},\label{DD1}\\
\delta\omega^{ab}  &  =\mathrm{D}_{\omega}\varepsilon^{ab},\label{DD2}\\
\delta k^{ab}  &  =\mathrm{D}_{\omega}\epsilon^{ab}+e^{a}\varepsilon^{b}%
-e^{b}\varepsilon^{a}-\varepsilon_{\text{ \ }c}^{a}k^{cb}+\varepsilon_{\text{
\ }c}^{b}k^{ca},\label{DD3}\\
\delta b^{a}  &  =\mathrm{D}_{\omega}\bar{\varepsilon}^{a}-\bar{\varepsilon
}_{\text{ \ }c}^{a}e^{c}+b_{\text{ \ }c}^{a}\varepsilon^{c}-\varepsilon
_{\text{ \ }c}^{a}b^{c},\label{DD4}\\
\delta b^{ab}  &  =\mathrm{D}_{\omega}\bar{\varepsilon}^{ab}-\varepsilon
_{\text{ \ }c}^{a}b^{cb}-\varepsilon_{\text{ \ }c}^{b}b^{ca},\label{DD5}\\
\delta\beta^{ab}  &  =\frac{1}{2}\mathrm{D}_{\omega}\bar{\epsilon}^{ab}%
+e^{a}\bar{\varepsilon}^{b}-\bar{\varepsilon}_{\text{ \ }c}^{a}k^{cb}%
-\varepsilon^{a}b^{b}-\varepsilon_{\text{ \ }c}^{a}B^{cb}+b_{\text{ \ }c}%
^{a}\epsilon^{cb}-k_{\text{ \ }e}^{a}k_{\text{ \ }g}^{e}\varepsilon^{g}%
e^{b}\nonumber\\
&  +k_{\text{ \ }e}^{a}k_{\text{ \ }h}^{e}e^{h}\varepsilon^{b}-2k^{ab}%
k_{\text{ \ }g}^{f}e_{f}\varepsilon^{g}-\epsilon_{\text{ \ }e}^{a}k_{\text{
\ }g}^{e}e^{g}e^{b}+\epsilon^{ab}k_{gh}e^{g}e^{h}-k^{ab}\epsilon_{gh}%
e^{g}e^{h}\nonumber\\
&  +k_{\text{ \ }e}^{a}\epsilon_{\text{ \ }g}^{e}e^{g}e^{b}-(a\leftrightarrow
b). \label{DD6}%
\end{align}
Notice that the information about the cocycle structure constants is contained
in the gauge variation of the three-form $\beta^{ab}$. This feature will be
inherited by the large gauge transformation and, as a consequence, an action
principle invariant under the enhanced symmetry of Maxwell-FDA must contain a
dynamic term for $\beta^{ab}$. Such requirement represents the main obstacle
when writing down a Chern--Simons action for Maxwell-FDA, since there is no
known invariant tensor involving that sector of the algebra.

Let us now consider a large gauge transformation in terms the sum of eq.
(\ref{fdalarge}). In this case, we choose the Lie subalgebra of Maxwell-FDA
spanned by $\left\{  J_{ab},Z_{cd}\right\}  $, and denoted by $\mathcal{L}%
_{\text{\textrm{Maxwell}}}$ as stability subalgebra. Algebras of this type can
be constructed as expansions of the Lorentz algebra and have been considered
in the construction of generalized Lovelock and MacDowell--Mansouri gravity
action principles in refs.
\cite{Concha:2013uhq,Concha:2014vka,Concha:2014zsa,Concha:2016tms}. The
multiplet is therefore chosen as
\begin{equation}
\xi\in\frac{\text{\textrm{Maxwell-FDA}}}{\mathcal{L}_{\text{\textrm{Maxwell}}%
}},
\end{equation}
to whose components we denote
\begin{equation}
\xi=\xi^{a}P_{a}+\zeta^{a}\bar{P}_{a}+\frac{1}{2}\zeta^{ab}\bar{J}_{ab}%
+\frac{1}{2}\kappa^{ab}\bar{Z}_{ab}.
\end{equation}
Since we do not consider a parametrization for the Lorentz-type subalgebra,
the first term $\Delta\mu$ in eq. (\ref{fdalarge}) has the same functional
form exhibited in eqs. (\ref{DD1})-(\ref{DD6}) but considering
non-infinitesimal parameters and setting $\varepsilon^{ab}=0$. By directly
calculating each following term, we find that the sum in eq. (\ref{fdalarge})
is truncated. Thus, we introduce the non-linear gauge field $\mu^{\xi}$ in
terms of $\mu$ and the coset space coordinates $\xi$ as%
\begin{equation}
\mu^{\xi}=V^{a}P_{a}+\frac{1}{2}W^{ab}J_{ab}+\frac{1}{2}K^{ab}Z_{ab}+B^{a}%
\bar{P}_{a}+\frac{1}{2}B^{ab}\bar{J}_{ab}+\frac{1}{2}\tilde{B}^{ab}\bar
{Z}_{ab},
\end{equation}
with%
\begin{align}
V^{a}  &  =e^{a}+\mathrm{D}_{\omega}\xi^{a},\\
W^{ab}  &  =\omega^{ab},\\
K^{ab}  &  =k^{ab}+e^{a}\xi^{b}-e^{b}\xi^{a}-\frac{1}{2}\xi^{a}\mathrm{D}%
_{\omega}\xi^{b}+\frac{1}{2}\xi^{b}\mathrm{D}_{\omega}\xi^{a},\\
B^{a}  &  =b^{a}+\mathrm{D}_{\omega}\zeta^{a}+b_{\text{ \ }c}^{a}\xi^{c}%
-\zeta_{\text{ \ }c}^{a}e^{c}+\frac{1}{2}\mathrm{D}_{\omega}\zeta^{ab}\xi
_{b}-\frac{1}{2}\zeta_{\text{ \ }b}^{a}\mathrm{D}_{\omega}\xi^{b},\\
B^{ab}  &  =b^{ab}+\mathrm{D}_{\omega}\zeta^{ab},\\
\tilde{B}^{ab}  &  =\beta^{ab}+\mathrm{D}_{\omega}\kappa^{ab}+\left(
e^{a}\zeta^{b}-\xi^{a}b^{b}-\zeta_{\text{ \ }c}^{a}k^{cb}+\frac{1}%
{2}\mathrm{D}_{\omega}\xi^{a}\zeta^{b}+\frac{1}{2}\zeta_{\text{ \ }c}^{a}%
\xi^{c}e^{b}-\frac{1}{2}\zeta_{\text{ \ }c}^{a}\xi^{b}e^{c}\right. \nonumber\\
&  +k^{ab}k_{cd}\xi^{c}\mathrm{D}_{\omega}\xi^{d}-k_{\text{ \ }c}^{a}k_{\text{
\ }d}^{c}\xi^{d}e^{b}+k_{\text{ \ }c}^{a}k_{\text{ \ }d}^{c}e^{d}\xi
^{b}+2k^{ab}k_{cd}\xi^{c}e^{d}-2\xi^{a}e^{b}k_{cd}\xi^{c}e^{d}\nonumber\\
&  -\frac{1}{2}k_{\text{ \ }d}^{a}k_{\text{ \ }e}^{d}\xi^{e}\mathrm{D}%
_{\omega}\xi^{b}+\frac{1}{2}k_{\text{ \ }c}^{a}k_{\text{ \ }d}^{c}%
\mathrm{D}_{\omega}\xi^{d}\xi^{b}+\frac{1}{2}k_{\text{ \ }c}^{a}\xi^{c}%
e_{d}\xi^{d}e^{b}-\frac{1}{2}\xi^{2}k_{\text{ \ }c}^{a}e^{c}e^{b}+\frac{1}%
{2}k_{\text{ \ }c}^{a}e^{c}\xi_{d}e^{d}\xi^{b}\nonumber\\
&  +\frac{1}{6}\xi^{a}\zeta_{\text{ \ }c}^{b}\mathrm{D}_{\omega}\xi^{c}%
+\frac{1}{6}\zeta_{\text{ \ }c}^{a}\xi^{c}\mathrm{D}_{\omega}\xi^{b}-\frac
{1}{6}\zeta_{\text{ \ }c}^{a}\mathrm{D}_{\omega}\xi^{c}\xi^{b}+\frac{1}{6}%
\xi^{a}k_{cd}\xi^{c}\mathrm{D}_{\omega}\xi^{d}e^{b}\nonumber\\
&  -\frac{1}{3}\xi^{2}k^{ab}\mathrm{D}_{\omega}\xi_{c}e^{c}+\frac{1}{3}%
k^{ab}\xi_{c}\mathrm{D}_{\omega}\xi^{c}\xi_{d}e^{d}+\frac{1}{6}k_{\text{ \ }%
c}^{a}\xi^{c}\xi_{d}\mathrm{D}_{\omega}\xi^{d}e^{b}-\frac{1}{6}\xi
^{2}k_{\text{ \ }c}^{a}\mathrm{D}_{\omega}\xi^{c}e^{b}\nonumber\\
&  \left.  -\frac{1}{6}k_{\text{ \ }c}^{a}\xi^{c}\mathrm{D}_{\omega}\xi
^{d}e_{d}\xi^{b}+\frac{1}{6}k_{\text{ \ }c}^{a}\mathrm{D}_{\omega}\xi^{c}%
\xi_{d}e^{d}\xi^{b}-\frac{5}{6}\xi^{a}\mathrm{D}_{\omega}\xi^{b}k_{cd}\xi
^{c}e^{d}-\left(  a\leftrightarrow b\right)  \right)  .
\end{align}
with $\xi^{2}=\xi^{a}\xi_{a}$. As before, the infinitesimal gauge
transformations of the components of $\mu$ are already known and given by eqs.
(\ref{DD1})-(\ref{DD6}). However, in order to complete the non-linear
realization of Maxwell-FDA, we must specify the transformation law of the
multiplet, which in turns, fix the transformation law of $\mu^{\xi}$. These
variations are obtained from eqs. (\ref{fdalinear}) and (\ref{hh}) and can be
found in appendix \ref{app2}.

By considering the Maxwell-FDA multiplet and the original gauge field $\mu$,
it is possible to write down invariant action principles in arbitrary
dimension. For this purpose, we first introduce the curvature associated to
the non-linear gauge field $\mu^{\xi}$%
\begin{equation}
R^{\xi}=\mathrm{d}\mu^{\xi}+\frac{1}{2}\left[  \mu^{\xi},\mu^{\xi}\right]
_{2}+\frac{1}{\left(  p+1\right)  !}\left[  \left(  \mu^{\xi}\right)
^{p+1}\right]  _{p+1}.
\end{equation}
In the chosen basis, we denote its standard and extended components as
$R^{\xi}=\left(  \mathcal{R}^{A},\mathcal{H}^{A}\right)  $, with%
\begin{align}
\mathcal{R}^{A}  &  =\left(  \mathcal{T}^{a},R^{ab},\mathcal{F}^{ab}\right)
,\\
\mathcal{H}^{A}  &  =\left(  \mathcal{H}^{a},\mathcal{H}^{ab},\mathcal{G}%
^{ab}\right)  .
\end{align}
Notice that, since $\omega^{ab}$ does not transform, the Lorentz curvature
remains the same than in eq. (\ref{mfda1b}). By using the non-linear gauge
field and its corresponding gauge curvature as building blocks, we consider
the following modification of the six-dimensional Einstein--Hilbert action
principle%
\begin{equation}
I^{\mathrm{Maxwell}\text{\textrm{-}}\mathrm{FDA}}=\kappa\int~\left(
\epsilon_{abcdef}R^{ab}V^{c}V^{d}V^{e}V^{f}+\mathcal{G}^{ab}V_{a}V_{b}\right)
, \label{action2}%
\end{equation}
where $\kappa$ is a dimensional constant. It is important to point out that we
include the components of the curvature $\mathcal{G}^{ab}$, associated to the
three-forms $\tilde{B}^{ab}$. This ensures the presence of the four-cocycle in
the Lagrangian, equations of motion and symmetry transformations of the action
principle. Moreover, the cocycle structure constants are also present in the
gauge transformation law of the components $\kappa^{ab}$ of the Maxwell-FDA
multiplet (see eqs. (\ref{kappa1}) and (\ref{kappa2}) in appendix \ref{app2}).
Such components are included in the definition of the three-form gauge field
$\tilde{B}^{ab}$ and consequently in $\mathcal{G}^{ab}$.

In order to obtain spacetime geometry, both the three-forms and the one-forms
$K^{ab}$ are interpreted as matter fields while $\omega^{ab}$ and $V^{a}$ as
the spin connection and vielbein of a gravity theory. Therefore, as we did in
section 4.1, we split the Lagrangian density inside the integral of eq.
(\ref{action2}) in terms of two contributions that we identify as pure gravity
and matter Lagrangian densities:%
\begin{equation}
I^{\mathrm{Maxwell}\text{\textrm{-}}\mathrm{FDA}}=\kappa\int\left(
L_{\text{\textrm{G}}}+L_{\text{\textrm{M}}}\right)  .
\end{equation}
Here, $L_{\text{\textrm{G}}}$ corresponds to the six-dimensional
Einstein--Hilbert Lagrangian, while $L_{\text{\textrm{M}}}$ is the remaining
contribution depending on the three-forms. We also introduce an
energy-momentum form $\tau_{a}$ and a spin form $\sigma_{ab}$ in terms of the
functional variations of $L_{\text{\textrm{M}}}$, according to eqs.
(\ref{emtensor}) and (\ref{spin}) respectively. In this case, the resulting
equations of motion are%
\begin{align}
\varepsilon_{a}  &  \equiv\frac{\delta L_{\text{\textrm{G}}}}{\delta V^{a}%
}=-4\epsilon_{abcdef}R^{bc}V^{d}V^{e}V^{f}=-\ast\tau_{a},\label{ep1}\\
\varepsilon_{ab}  &  \equiv\frac{\delta L_{\text{\textrm{G}}}}{\delta
\omega^{ab}}=4\epsilon_{abcdef}\mathcal{T}^{c}V^{d}V^{e}V^{f}=-\ast\sigma
_{ab}, \label{ep2}%
\end{align}
where%
\begin{align}
\ast\tau_{a}  &  =-2\mathrm{D}_{\omega}\tilde{B}_{ab}V^{b}-4K_{a}^{\text{
\ }c}B_{cb}V^{b}+2KK_{ab}V^{b},\label{em2}\\
\ast\sigma_{ab}  &  =\tilde{B}_{b}^{\text{ \ }c}V_{c}V_{a}-\tilde{B}%
_{a}^{\text{ \ }c}V_{c}V_{b}, \label{spin2}%
\end{align}
with $K=K_{ab}V^{a}V^{b}$. As it happens in the Poincare-FDA case, the
presence of $p$-forms leads to non-vanishing torsion in eq. (\ref{ep2}).
However, in contrast to that case, considering absence of higher-degree forms
does not lead to $\varepsilon_{ab}=0$. In such case, the torsion is indeed
vanishing but the Einstein equation is modified by the presence of the cocycle
in the third term in the r.h.s. of eq. (\ref{spin2}), which has a topological
origin. Notice that the cocycle was introduced in the gauge algebra by virtue
of the presence of the higher-degree forms but, once the gauge algebra is
modified, the resulting gravity theory shows a diferent behavior even in
absence of these higher fields.

It is noteworthy that, although FDAs initially emerged in the study of
high-dimensional supergravity, the existence of Chevalley--Eilenberg
cohomology classes is not limited to superalgebras. This feature enables the
formulation of gauge theories involving $p$-forms within the framework of
purely bosonic gravity. For instance, ref. \cite{Bonanos:2008kr} explores
cohomology classes for bosonic algebras and their role in extending
Poincar\'{e} and Galilei symmetries. Similarly, ref. \cite{Camarero:2017yka}
introduces a bosonic Chern--Simons gravity theory with $p$-forms, constructed
by gauging FDAs, as the bosonic sector of a $D=11$ supergravity theory. The
bosonic FDA extension of the Maxwell algebra, alongside its corresponding
action principle, can likewise be interpreted as the bosonic sector of a
supergravity theory coupling three-forms. This interpretation also holds for
the Poincar\'{e} counterparts discussed in Section 4.1.

\section{Eleven-dimensional supergravity}

In this section we consider the gauging of the FDA of $D=11\ $supergravity in
the SW-GN formalism. This theory, first introduced in second order formalism
by Cremmer, Julia and Scherk (CJS) in ref. \cite{Cremmer:1978km}, emerges as
the low-energy limit of the M-theory and presents a field content composed by
the elfbein $e^{a}$, the spin connection $\omega^{ab}$, the gravitino field
$\psi$ and a three-form gauge field $A$ carrying no algebraic index. Later
studies shown a first order formalism for its action principle was possible,
where the presence of torsion and supertorsion was allowed as a consequence of
independent degrees of freedom for the spin connection \cite{Julia:1999tk}.
The first three mentioned fields are one-forms, which suggest that the
symmetry algebra underlying the theory must have the $D=11$ Poincar\'{e}
superalgebra as a subalgebra. Then, these fields can be considered as the
gauge fields associated to the translation, Lorentz rotations and
supersymmetry generators respectvely, as it is usually done when gauging Lie
superalgebras. However, the presence of the three-form $A$ makes this gauging
impossible. This means that the symmetries underlying the theory must be
described by a generalized structure that naturally includes higher-degree
forms in their gauging, as it is the case of FDAs. The FDA of $D=11$
supergravity was introduced in ref. \cite{DAuria:1982uck} where a geometric
formulation of the CJS supergravity was presented. This FDA is defined by
means of the following Maurer--Cartan equations:%
\begin{align}
R^{a}  &  =\mathrm{d}e^{a}-\omega_{\text{ \ }b}^{a}e^{b}-\frac{i}{2}\bar{\psi
}\Gamma^{a}\psi=0,\label{fda1}\\
R^{ab}  &  =\mathrm{d}\omega^{ab}-\omega_{\text{ \ }c}^{a}\omega
^{cb}=0,\label{fda2}\\
\rho &  =\mathrm{d}\psi-\frac{1}{4}\omega^{ab}\Gamma_{ab}\psi=0,\label{fda3}\\
F  &  =\mathrm{d}A-\frac{1}{2}\bar{\psi}\Gamma_{ab}\psi e^{a}e^{b}=0,
\label{fda4}%
\end{align}
where $\psi,\rho$ are Majorana spinors and the Lorentz indices are raised and
lowered with the Minkowski metric $\eta_{ab}=\mathrm{diag}\left(
+,-,\ldots,-\right)  $. It is important to point out the presence of a
non-trivial four-cocycle in eq. (\ref{fda4}), representative of a cohomology
class, which makes impossible the writing of the three-form $A$ in terms of
the one-forms of the Lie subalgebra $e^{a}$, $\omega^{ab}$ and $\psi$. For
details on the non-triviality of this cocycle and higher-dimensional ones, see
refs. \cite{DAuria:1982uck,Bandos:2004ym,Ravera:2021sly}.

\subsection{Infinitesimal gauge transformation}

The FDA of $D=11$ supergravity can be treated in the terms of eqs. (\ref{mu})
and (\ref{B}). From eqs. (\ref{fda1})-(\ref{fda3}) we can see that the
one-form $\mu^{A}$ is composed by $\left(  e^{a},\omega^{ab},\psi\right)  $.
Thus, the Maurer--Cartan equation (\ref{mu})~describes the Poincar\'{e}
superalgebra. Moreover, contrary to the FDAs presented in sections 4.1 and
4.2, the three-form $A$ is not introduced in the adjoint representation of the
Poincar\'{e} superalgebra since it carries no algebraic index. This means that
the algebraic index $i$ of eq. (\ref{B}) takes one single value $\mu^{i}\equiv
A$. We can now equivalently write this FDA1 as an $L_{\infty}$ algebra spanned
by the vectors $\left\{  t_{A},t_{i}\right\}  $. We denote the components of
the dual vectors $t_{A}$ and $t_{i}$ as follows%
\begin{equation}
t_{A}=\left(  P_{a},J_{ab},Q_{\alpha}\right)  ,\text{ \ \ \ \ }t_{i}=\left\{
t\right\}  .
\end{equation}
Note that this is a FDA1 in which the generalized structure constants of the
type $C_{Aj}^{i}$ vanish, but those associated to the coclycle $C_{ABCD}^{i}$
do not. From eqs. (\ref{fda1})-(\ref{fda4}) it is possible to extract the
information regarding the non-vanishing structure constants and codify them
into a $L_{\infty}$ algebra. Thus, eqs. (\ref{fda1A})-(\ref{fda1C}) take the
form
\begin{align}
\left[  J_{ab},J_{cd}\right]   &  =\eta_{ac}J_{bd}+\eta_{bd}J_{ac}-\eta
_{bc}J_{ad}-\eta_{ad}J_{bc},\\
\left[  J_{ab},P_{c}\right]   &  =\eta_{ac}P_{b}-\eta_{bc}P_{a},\\
\left\{  Q_{\alpha},Q_{\beta}\right\}   &  =-i \left(  C\Gamma
^{a}\right)  _{\alpha\beta}P_{a},\\
\left[  J_{ab},Q_{\gamma}\right]   &  =-\frac{1}{2}\left(  \Gamma_{ab}\right)
_{\text{ \ }\gamma}^{\alpha}Q_{\alpha},\\
\left[  Q_{\alpha},Q_{\beta},P_{a},P_{b}\right]   &  =-12\left(  C\Gamma
_{ab}\right)  _{\alpha\beta}t.
\end{align}
In the context of FDA1 gauge theory, the gauge field can be spanned as a
linear combination of the dual $L_{\infty}$ algebra basis vectors, as follows%
\begin{equation}
\mu=e^{a}P_{a}+\frac{1}{2}\omega^{ab}J_{ab}+\psi^{\alpha}Q_{\alpha}+At.
\end{equation}
Let us now consider the following infinitesimal gauge parameter%
\begin{equation}
\varepsilon=\varepsilon^{a}P_{a}+\frac{1}{2}\varepsilon^{ab}J_{ab}%
+\theta^{\alpha}Q_{\alpha}+\epsilon t,
\end{equation}
where $\varepsilon^{a}$ and $\varepsilon^{ab}$ are zero-forms, $\theta
^{\alpha}$ is a zero-form Majorana spinor, and $\bar{\varepsilon}$ is a
two-form. The infinitesimal gauge variations of the gauge fields are given by%
\begin{align}
\delta e^{a}  &  =\mathrm{d}\varepsilon^{a}-\omega_{\text{ \ }b}%
^{a}\varepsilon^{b}+\varepsilon_{\text{ \ }b}^{a}e^{b}-i\psi^{\alpha}\left(
C\Gamma^{a}\right)  _{\alpha\beta}\theta^{\beta},\label{svar1}\\
\delta\omega^{ab}  &  =\mathrm{d}\varepsilon^{ab}-\omega_{\text{ \ }c}%
^{a}\varepsilon^{cb}+\omega_{\text{ \ }c}^{b}\varepsilon^{ca},\label{svar2}\\
\delta\psi^{\alpha}  &  =\mathrm{d}\theta^{\alpha}-\frac{1}{4}\omega
^{ab}\left(  \Gamma_{ab}\right)  _{\text{ \ }\beta}^{\alpha}\theta^{\beta
}+\frac{1}{4}\varepsilon^{ab}\left(  \Gamma_{ab}\right)  _{\text{ \ }\beta
}^{\alpha}\psi^{\beta},\label{svar3}\\
\delta A  &  =\mathrm{d}\epsilon+\psi^{\alpha}\left(  C\Gamma_{ab}\right)
_{\alpha\beta}\psi^{\beta}\varepsilon^{a}e^{b}+\theta^{\alpha}\left(
C\Gamma_{ab}\right)  _{\alpha\beta}\psi^{\beta}e^{a}e^{b}. \label{svar4}%
\end{align}

\subsection{Non-linear realization}

Let us now consider the construction of a non-linear gauge field. To this end,
we identify the Lorentz subalgebra as stability subalgebra $\mathfrak{h}%
=\mathfrak{so}\left(  1,10\right)  $. Moreover, we introduce the following
multiplet lyng on the coset space FDA1$/\mathfrak{h}$%
\begin{equation}
\xi=\xi^{a}P_{a}+\chi^{\alpha}Q_{\alpha}+\zeta t.
\end{equation}
We now introduce the non-linear gauge field $\mu^{\xi}$ in terms of the
components of $\mu$ and the coset space coordinates $\xi$. We denote it as%
\begin{equation}
\mu^{\xi}=V^{a}P_{a}+\frac{1}{2}W^{ab}J_{ab}+\Psi^{\alpha}Q_{\alpha
}+\mathcal{A}t,
\end{equation}
where each component is given by
\begin{align}
V^{a}  &  =e^{a}+\mathrm{D}_{\omega}\xi^{a}-i\left(  \bar{\psi}\Gamma^{a}%
\chi\right)  +\frac{i}{2}\left(  \bar{\chi}\Gamma^{a}\mathrm{D}_{\omega}%
\chi\right)  ,\\
W^{ab}  &  =\omega^{ab},\\
\Psi^{\alpha}  &  =\bar{\psi}^{\alpha}+\mathrm{D}_{\omega}\bar{\chi}^{\alpha
},\\
\mathcal{A}  &  =A+\mathrm{d}\zeta+\left(  \bar{\psi}\Gamma_{ab}\psi\right)
\xi^{a}e^{b}+\left(  \bar{\chi}\Gamma_{ab}\psi\right)  e^{a}e^{b}+\frac{1}%
{2}\xi^{a}\mathrm{D}_{\omega}\xi^{b}\left(  \bar{\psi}\Gamma_{ab}\psi\right)
-\frac{i}{2}\xi^{a}\left(  \bar{\psi}\Gamma^{b}\chi\right)  \left(  \bar{\psi
}\Gamma_{ab}\psi\right) \nonumber\\
&  +\left(  \mathrm{D}_{\omega}\bar{\chi}\Gamma_{ab}\psi\right)  \xi^{a}%
e^{b}-\mathrm{D}_{\omega}\xi^{a}\left(  \bar{\chi}\Gamma_{ab}\psi\right)
e^{b}+i\left(  \bar{\psi}\Gamma^{a}\chi\right)  \left(  \bar{\chi}\Gamma
_{ab}\psi\right)  e^{b}+\frac{1}{2}\left(  \bar{\chi}\Gamma_{ab}%
\mathrm{D}_{\omega}\chi\right)  e^{a}e^{b}\nonumber\\
&  +\frac{i}{6}\xi^{a}\left(  \bar{\chi}\Gamma^{b}\mathrm{D}_{\omega}%
\chi\right)  \left(  \bar{\psi}\Gamma_{ab}\psi\right)  +\frac{i}{3}\left(
\bar{\chi}\Gamma_{ab}\psi\right)  \left(  \bar{\chi}\Gamma^{a}\mathrm{D}%
_{\omega}\chi\right)  e^{b}.
\end{align}
We now identify $V^{a}$ as the elfbein of $D=11$ supergravity, $\Psi$ as
gravitino one-form, and $\mathcal{A}$ as the independent three-form gauge
field of CJS supergravity. $\omega^{ab}$ holds its role as spin connection
one-form. The infinitesimal gauge transformation law of these non-linear gauge
fields are fixed by eqs. (\ref{svar1})-(\ref{svar4}), in addition to the
transformation law of the FDA multiplet (see appendix \ref{app3}). This
completes the non-linear realization of the FDA of $D=11$ supergravity. It is
now possible to write a gauge invariant action principle by considering a
Lorentz invariant Lagrangian and then writing the action in terms of $\mu
^{\xi}$ in replacement of $\mu$. For this purpose, we consider the following
functional%
\begin{align}
I\left[  \mu^{\xi}\right]   &  =\int-\frac{1}{4}R^{ab}\Sigma_{ab}+\frac{i}%
{2}\bar{\Psi}\Gamma^{\left(  8\right)  }\mathrm{D}_{\omega}\Psi+\frac{i}%
{8}\left(  \mathrm{D}_{\omega}V^{a}-\frac{i}{4}\bar{\Psi}\Gamma^{a}%
\Psi\right)  V_{a}\bar{\Psi}\Gamma_{\left(  6\right)  }\Psi-\frac{1}%
{2}\mathcal{F}\ast\mathcal{F}\nonumber\\
&  +\left(  \ast\mathcal{F}+\frac{i}{4}\bar{\Psi}\Gamma_{\left(  5\right)
}\Psi\right)  \left(  \mathrm{d}\mathcal{A}-\frac{i}{4}\bar{\Psi}%
\Gamma_{\left(  2\right)  }\Psi\right)  -\frac{1}{32}\left(  \bar{\Psi}%
\Gamma_{\left(  2\right)  }\Psi\right)  \left(  \bar{\Psi}\Gamma_{\left(
5\right)  }\Psi\right)  -\frac{1}{3}\mathrm{d}\mathcal{A}\mathrm{d}%
\mathcal{A}\mathrm{d}\mathcal{A}, \label{action11}%
\end{align}
where $\mathcal{F}$ is the curvature four-form associated to $\mathcal{A}$ by
means of eq. (\ref{fda3}) in presence of non-linear gauge fields, and where we
consider the following notation
\begin{equation}
\Sigma_{ab}=\frac{1}{9!}\epsilon_{aba_{1}\cdots a_{9}}V^{a_{1}}\ldots
V^{a_{9}},\text{ \ \ \ \ }\Gamma_{\left(  n\right)  }=\frac{1}{n!}%
\Gamma_{a_{1}\ldots a_{n}}V^{a_{1}}\ldots V^{a_{n}}.
\end{equation}
The action (\ref{action11}) was proposed in \cite{Julia:1999tk} as the first
order formalism version of the original action introduced in ref.
\cite{Cremmer:1978km}. However, since we have chosen the components of the
non-linear gauge field as the fundamental fields of the theory, the action
remains invariant under the transformations of the FDA (\ref{fda1}%
)-(\ref{fda4}). This permits to formulate the action (\ref{action11}) as a
genuine gauge theory of a FDA without decomposing the three-form as a
combination of products of one-forms.

\section{Concluding Remarks}

In this article we have considered an extension of the SW-GN formalism in the
context of a particular family of FDAs known as FDA1. In Lie group theory,
such formalism requires a set of zero-form scalar fields which is
mathematically introduced as the parameter of a large gauge transformation. In
order to introduce these transformations in FDA gauge theories, we have made
use of the dual vectors of the associated $L_{\infty}$ algebras to introduce a
compact notation in which the FDA gauge parameters and fields can be written
as single matemathical ojects. We have shown that, in absence of a cocycle in
the construction of a FDA1, the generalized structure constants corresponding
to the bilinear products of the FDA, satisfy the standard Jacobi identity.
Consequently, for this kind of FDA, large gauge transformations can be
introduced by applying the standard SW-GN formalism, which makes possible to
write down non-linear realizations and invariant action principles for gauge
theories with $p$-forms. The use the mentioned dual basis has allowed us to
write vectors and connections evaluated in FDAs as linear combinations of
vectors, just as it is usually done in the study of Lie algebras. This
approach has been particularly useful when calculating FDA-evaluated gauge
transformations and curvatures whose components are given by differential
forms of different degrees. Moreover, we have extended the mentioned study to
the case of an arbitrary FDA1, i.e., an algebra carrying one $p$-form
extension that includes a representative of a non-trivial Chevalley--Eilemberg
cohomology class in its Maurer--Cartan equation. In this case, we have
introduced the large gauge transformations without make use of an
exponentiation map but by analytically extending the gauge transformations of
Lie algebras in a way that couples the higher-degree differential forms and
the cocycle. We have studied two bosonic cases, namely, two FDA1 extensions of
the Poincar\'{e} and Maxwell Lie algebras in arbitrary dimensions, carrying
non-trivial cocycles, each one representative of a different cohomology class.
For these algebras, we have found non-linear realizations and proposed gauge
invariant action principles that extend the Einstein--Hilbert action by
including three-forms in the adjoint representation of mentioned Lie algebras.
Of particular interest is the case of Maxwell-FDA, where the presence of the
cocycle in the equations of motion modifies the standard theory even if we set
the higher-degree forms as vanishing. Moreover, we have considered an
application of the formalism in the context of higher-dimensional
supergravity. Specifically, we have found the $L_{\infty}$ algebra dual to the
FDA of the eleven-dimensional CJS supergravity. Then, we have gauged the
algebra by applying the extension of the SW-GN formalism, which has allowed us
to formulate this theory as a genuine gauge theory of such FDA. It is
noteworthy that, in ref. \cite{DAuria:1982uck}, it was shown that the
underlying symmetry of $D=11$ supergravity can be alternatively described by a
Lie superalgebra. This is carried out by considering that the three-form $A$
can be decomposed as a sum of products of the one-forms of the Poincar\'{e}
superalgebra $\left(  e^{a},\omega^{ab},\psi\right)  $, in addition to two new
bosonic one-forms $B^{ab}=B^{\left[  ab\right]  }$ and $B^{abcde}=B^{\left[
abcde\right]  }$, and a fermionic one-form $\eta$. Consequently, the dual
basis of vectors is extended by the inclussion of two bosonic generators and a
fermionic one. This structure is known as the D'Auria--Fr\'{e} superalgebra.
It is important to point out that the cocycle of the extended Maurer--Cartan
equation (\ref{fda4}) is non-trivial, i.e., it belongs to a
Chevalley--Eilenberg cohomology class. Its non triviality makes impossible the
writing of the three-form $A$ in terms of the one-forms of the Lie subalgebra.
This problem was solved by means of the inclussion of the new one-forms (and
generators), which extend the Poincar\'{e} superalgebra in a way that is
consistent with the structure equation of $A\,$. In this context, a similar
approach to the one presented in section 5 was carried out in ref.
\cite{Perez:2008zza}, were the standard SW-GN formalism was used to write down
the $D=11$ supergravity action as a gauge invariant action of the
D'Auria--Fr\'{e} superalgebra.

The presented extension of the SW-GN formalism motivates us to study several
aspects of Lie gauge theories that make use of large gauge transformations in
the context of FDAs. For example, it would be interesting to extend the
Stora--Zumino descent chain to the case of FDAs and to study the topological
quantities and generalized gauge anomalies that emerge from it
\cite{Zumino:1983ew,Zumino:1983rz,Manes:1985df}. Additionally, it would be
possible to study gauged Wess--Zumino--Witten models for FDAs, which would
allow the construction of dimensionally reduced gauge theories coupling
higher-degree forms without breaking the gauge invariance
\cite{Anabalon:2006fj,Anabalon:2007dr,Anabalon:2008hi,Mora:2011sz}. Moreover,
it would be also interesting to study the existence of supersymmetric
extensions of the Poincar\'{e} and Maxwell cocycles of section 4. This could
lead to supersymmetric versions of the presented FDAs and further applications
in the framework of supergravity theories with $p$-forms.

\section*{Acknowledgements}

The author would like to thank Patrick Concha, Evelyn Rodr\'{\i}guez and
Patricio Salgado for enlightening discussions. The author is grateful to
Universidad de Tarapac\'{a}, Chile, for the total support of this work.

\appendix

\section{Variations of the FDA multiplets}

Under an infinitesimal gauge transformation with gauge parameter evaluated on
the stability algebra or the coset space, the components of $\xi$ transform
according eqs. (\ref{fdalinear}) and (\ref{hh}) respectively.

\subsection{Poincar\'{e}-FDA}

\label{app1}

Let us first consider an infinitesimal parameter $\varepsilon$ evaluated in
the stability algebra (in this case, the Lorentz algebra), to whose components
we denote as%
\begin{equation}
\varepsilon=\frac{1}{2}\varepsilon^{ab}J_{ab}. \label{param1}%
\end{equation}
By plugging in eq. (\ref{param1}) into eq. (\ref{fdalinear}), we find the
following transformation rules for the zero-forms and two-forms of the
Poincar\'{e}-FDA multiplet:%
\begin{align}
\delta\xi^{a}  &  =-\varepsilon_{\text{ \ }c}^{a}\xi^{c},\\
\delta\zeta^{a}  &  =-\varepsilon_{\text{ \ }c}^{a}\zeta^{c}+3\xi
^{a}\varepsilon_{bc}\omega_{\text{ \ }e}^{b}\omega^{ec},\\
\delta\zeta^{ab}  &  =-\varepsilon_{\text{ \ }c}^{a}\zeta^{cb}+\varepsilon
_{\text{ \ }c}^{b}\zeta^{ca}.
\end{align}
On the other hand, by considering a parameter evaluated in the coset space,
given by%
\begin{equation}
\varepsilon=\varepsilon^{a}P_{a}+\bar{\varepsilon}^{a}\bar{P}_{a}+\frac{1}%
{2}\bar{\varepsilon}^{ab}\bar{J}_{ab},
\end{equation}
eq. (\ref{hh}) leads to the following transformation rule for the components
of the multiplet%
\begin{align}
\delta\xi^{a}  &  =-\varepsilon^{a},\\
\delta\zeta^{a}  &  =-\bar{\varepsilon}^{a}-\frac{1}{2}\bar{\varepsilon
}_{\text{ \ }c}^{a}\xi^{c}+\frac{1}{2}\zeta_{\text{ \ }c}^{a}\varepsilon
^{c},\\
\delta\zeta^{ab}  &  =-\bar{\varepsilon}^{ab}.
\end{align}

\subsection{Maxwell-FDA}

\label{app2}

Let us now consider an infinitesimal gauge transformation with gauge parameter
lying on the chosen stability algebra, which in this case is spanned by
$\left\{  J_{ab},Z_{ab}\right\}  $%
\begin{equation}
\varepsilon=\frac{1}{2}\varepsilon^{ab}J_{ab}+\frac{1}{2}\epsilon^{ab}Z_{ab}.
\end{equation}
In this case, eq. (\ref{fdalinear}) leads to the following transformation rule
for the components of the Maxwell-FDA multiplet%
\begin{align}
\delta\xi^{a}  &  =-\varepsilon_{\text{ \ }c}^{a}\xi^{c},\\
\delta\zeta^{a}  &  =-\varepsilon_{\text{ \ }c}^{a}\zeta^{c},\\
\delta\zeta^{ab}  &  =-\varepsilon_{\text{ \ }c}^{a}\zeta^{cb}+\varepsilon
_{\text{ \ }c}^{b}\zeta^{ca},\\
\delta\kappa^{ab}  &  =-\varepsilon_{\text{ \ }c}^{a}\kappa^{cb}+\zeta_{\text{
\ }c}^{a}\epsilon^{cb}+\epsilon_{\text{ \ }c}^{a}k_{\text{ \ }d}^{c}\xi
^{d}e^{b}-\epsilon_{\text{ \ }c}^{a}k_{\text{ \ }d}^{c}e^{d}\xi^{b}%
-2\epsilon^{ab}k_{cd}\xi^{c}e^{d}\nonumber\\
&  -k_{\text{ \ }c}^{a}\epsilon_{\text{~\ }d}^{c}\xi^{d}e^{b}+2k^{ab}%
\epsilon_{cd}\xi^{c}e^{d}+k_{\text{ \ }c}^{a}\epsilon_{\text{ \ }d}^{c}%
e^{d}\xi^{b}-(a\leftrightarrow b). \label{kappa1}%
\end{align}
Furthermore, by considering a gauge parameter lying in the coset space between
Maxwell-FDA and the mentioned stability algebra, to whose components we denote
as
\begin{equation}
\varepsilon=\varepsilon^{a}P_{a}+\bar{\varepsilon}^{a}P_{a}+\frac{1}{2}%
\bar{\varepsilon}^{ab}\bar{J}_{ab}+\frac{1}{2}\bar{\epsilon}^{ab}\bar{Z}_{ab},
\end{equation}
eq. (\ref{hh}) implies the following transfomation rule for the components of
$\xi\,$:%
\begin{align}
\delta\xi^{a}  &  =-\varepsilon^{a},\\
\delta\zeta^{a}  &  =-\bar{\varepsilon}^{a}+\frac{1}{2}\zeta_{\text{ \ }c}%
^{a}\varepsilon^{c}-\frac{1}{2}\bar{\varepsilon}_{\text{ \ }c}^{a}\xi^{c},\\
\delta\zeta^{ab}  &  =-\bar{\varepsilon}^{ab},\\
\delta\kappa^{ab}  &  =-\frac{1}{2}\bar{\epsilon}^{ab}+\frac{1}{4}k^{ac}%
k_{cd}\xi^{d}\varepsilon^{b}-\frac{1}{4}k^{ac}k_{cd}\varepsilon^{d}\xi
^{b}-\frac{1}{4}k^{bc}k_{cd}\xi^{d}\varepsilon^{a}\nonumber\\
&  +\frac{1}{4}k^{bc}k_{cd}\varepsilon^{d}\xi^{a}-k^{ab}k_{cd}\xi
^{c}\varepsilon^{d}-(a\leftrightarrow b). \label{kappa2}%
\end{align}

\subsection{The FDA of $D=11$ supergravity}

\label{app3}

In this case, the stability algebra is chosen as $\mathfrak{so}\left(
1,10\right)  $. We therefore consider an infinitesimal parameter given by%
\begin{equation}
\varepsilon=\frac{1}{2}\varepsilon^{ab}J_{ab}.
\end{equation}
In this case, eq. (\ref{fdalinear}) lead to the following transformation law
for the components of $\xi$:%
\begin{equation}
\delta\xi^{a}=-\varepsilon_{\text{ \ }c}^{a}\xi^{c},\text{ \ \ \ \ }\delta
\chi=\frac{1}{4}\varepsilon^{ab}\Gamma_{ab}\chi,\text{ \ \ \ \ }\delta\zeta=0.
\end{equation}
Furthermore, if we consider a gauge a parameter evaluated in the coset space,
given by%
\begin{equation}
\varepsilon=\varepsilon^{a}P_{a}+\theta^{\alpha}Q_{\alpha}+\epsilon t,
\end{equation}
we find that the components of the multiplet transform according to%
\begin{align}
\delta\xi^{a}  &  =-\varepsilon^{a}P_{a}-\frac{1}{2}i\bar{\chi}\Gamma
^{a}\theta,\\
\delta\chi^{\alpha}  &  =-\theta^{\alpha},\\
\delta\zeta &  =-\epsilon-\varepsilon^{a}e^{b}\bar{\chi}\Gamma_{ab}\psi
+\xi^{a}e^{b}\bar{\theta}\Gamma_{ab}\psi-\frac{1}{2}\xi^{a}\varepsilon^{b}%
\bar{\psi}\Gamma_{ab}\psi\nonumber\\
&  -\frac{i}{6}\left(  \bar{\chi}\Gamma^{a}\theta\right)  e^{b}\left(
\bar{\chi}\Gamma_{ab}\psi\right)  -\frac{i}{12}\xi^{a}\left(  \bar{\chi}%
\Gamma^{b}\theta\right)  \left(  \bar{\psi}\Gamma_{ab}\psi\right)  .
\end{align}

\bibliographystyle{utphys.bst}
 
\bibliography{RefsNL}

\end{document}